\documentclass[a4paper,10pt]{article}
\usepackage[utf8x]{inputenc}
\usepackage{amssymb, amsmath}
\usepackage{graphicx}
\usepackage{tabularx}
\usepackage{booktabs}
\usepackage{pdflscape}
\usepackage{lscape}
\usepackage{rotating}
\usepackage{float}
\usepackage{colortbl}
\usepackage{longtable} 
\usepackage{subcaption}
\usepackage[round]{natbib}
\usepackage{authblk}

\title{Risk diversification: a study of persistence with a filtered correlation-network approach}

\author[1]{Nicol\'o Musmeci} 
\author[2,3]{Tomaso Aste} 
\author[1]{T. Di Matteo}

\affil[1]{Department of Mathematics, King's College London, The Strand, London, WC2R 2LS}
\affil[2]{Department of Computer Science, UCL, Gower Street, London, WC1E 6BT, UK}
\affil[3]{Systemic Risk Centre, London School of Economics and Political Sciences, London, WC2A2AE, UK}

\begin{document}

\maketitle


\begin{abstract}
The evolution with time of the correlation structure of equity returns is studied by means of a filtered network approach investigating persistences and recurrences and their implications for risk diversification strategies. 
We build dynamically Planar Maximally Filtered Graphs  from the correlation structure over a rolling window and we study the persistence of the associated Directed Bubble Hierarchical Tree  (DBHT) clustering structure. 
We observe that the DBHT  clustering structure is quite stable during the early 2000' becoming gradually less persistent before the unfolding of the 2007-2008  crisis. 
The correlation structure eventually recovers persistence in the aftermath of the crisis settling up a new phase, distinct from the pre-cysts structure, where the market structure is less related to industrial sector activity.
Notably, we observe that -presently- the correlation structure is loosing again persistence indicating the building-up of another, different, phase. 
Such dynamical changes in persistence and their occurrence at the unfolding of financial crises rises concerns about the effectiveness of correlation-based portfolio management tools for risk diversification.
\end{abstract}

\section*{Introduction}
A way to reduce financial risk is diversifying investments taking positions in assets that are historically anti-correlated or uncorrelated reducing in this way the probability that all assets loose value at the same time. 
This is, for instance, the idea at the basis of the Capital Asset Pricing Model \citep{fama2004capital}. 
However, the applicability of these approaches relies on the implicit assumption that the relevant features of the correlation structure observed in the past have persistent significancy into the future. 
This is not always the case.

In order to characterise the correlation structure and quantify its persistence, in this paper we use a network filtering approach where the correlation matrix is mapped into a sparse graph that retains the relevant elements only. 
To this purpose we use the correlation filtered networks know as Planar Maximally Filtered Graph (PMFG) \citep{Tumminello05} and its associated clustering structure Directed Bubble Hierarchical Tree (DBHT) \citep{DBHT}. 
PMFG is a maximal planar graph that retains the largest correlations only. 
The DBHT is a hierarchical clustering that is  constructed making use of the separating properties of 3-cliques in planar graphs \citep{bubble}.

Since the seminal work by \cite{mantegna1}, network analysis on asset correlation has provided interesting insights for risk management and portfolio optimisation. 
It has been observed that the structure of such networks has significant relations with the industrial sectors classifications but also conveys important independent information \citep{mantegna1,musmeci_DBHT}.
It has been shown that this network structure \citep{market_mode} can be very robust against changes in the time horizon at which the assets returns are sampled (when the market mode dynamics is removed from the original correlations). 
This has been interpreted as an indication ``that correlations on short time scales might be used as a proxy for correlations on longer time horizons'' \citep{market_mode}. 
This however requires some degree of stationarity in the correlation structure.

Network filtering procedures have been found to improve sensibly the performance of portfolio optimisation methods. 
For instance, it has been shown in \citep{cluster_portfolio} that Markowitz optimisation gives better results on network-filtered correlation matrices than on unfiltered ones. 
In \cite{invest_periph} it has been reported that the peripheral position of nodes in PMFGs can be a criterion to select a well-diversified portfolio. 
This finding is consistent with what found for the Maximum Spanning Three (MST) in \cite{taxonomy_portfolio}, namely that the stocks selected by Markowitz method tend to be the ``leaves'' of the MST.

Network filtered correlations carry both local and global informations in their structure and the analysis of their temporal evolution can allow to better understand financial market evolution.
For instance, in \cite{pozzi_dyn_net}, it has been observed that stocks belonging to the same industrial sector tend to have similar values of centrality in the network topology and that this differentiation is quite persistent over time. 
In particular, they observed that Finance, Basic Materials and Capital Goods industrial sectors (Forbes classification) tend to be the located mostly in the central region of the network whereas Energy, Utilities and Health Care are located more in the peripheral region.
The preeminent role of the Financial sector is even stronger when correlation networks based on partial correlations are analysed \citep{partial_corr}.
Despite this overall robustness, a certain degree of non-stationarity has been also observed.
For instance, the Financial sector appears to loose centrality over the first decade of 2000's \citep{NJP10}. 
In \cite{buccheri} the authors found both a slow and a fast dynamics in correlation networks topology:
while the slow dynamics shows persistence over periods of at least 5 years,
 the time scale of the fast dynamics is of order of few months and it is linked to special exogenous and endogenous events like financial crises. 
For instance, in \cite{black_monday} it has been shown that sharp structural changes occurred in the graph topology during the 1987 Black Monday. 
Similar phenomena have been observed for correlation on Foreign Exchange (FX) data \citep{fx_crisis_mst}. 
In \cite{impact_events_mst} it has been demonstrated that structural changes on FX correlation data display different features depending on the type of event affecting the market: news that concern economic matters can trigger a prompt destabilising reaction, whereas there are periods of ``collective discovery'' where opinions appear to synchronise \citep{impact_events_mst}. 
  
 In this work we investigate the non-stationarity of correlation  quantifying how much, and in which way, the correlation structure changes over time.
 This is a particularly relevant topic because most of the portfolio optimisation tools rely on some stationarity or -at least- persistence  in the joint distribution of asset returns. 
 It is generally accepted in the literature that financial correlations are non-stationary. 
 For instance, in \cite{non_stationary_corr} it has been shown, by means of local Kolmogorov-Smirnov test on correlation pairs, how non-stationarity can affect sensitively the effectiveness of portfolio optimisation tools. 
 In this paper we discuss the degree of non-stationarity in the correlations at non-local level by using PMFG networks and the associated DBHT clustering and looking at the changes in the hierarchical and clustering structures. 
In this context persistence translates into a measure of similarity among communities in a network, for which network-theoretic tools should be used.  
PMFG-DBHT method has been recently applied to the study of financial data  \citep{musmeci_DBHT} showing that it is a powerful clustering tool that can outperform other traditional clustering methods such as Linkage and k-medoids in retrieving economic information.
 Moreover, the dynamical analyses have shown that  the clustering structure reveals peculiar patterns over the financial crisis showing an increasing dominant role of the market mode over the period 1997-2012, implying an increase of the non-diversifiable risk in the market. 
In this paper we take these analyses a step further looking at the dynamics of this clustering and its persistence. 

 The rest of the paper is organised as follows: in ``Correlation-based networks: an overview'' we summarise the main theoretical concepts underlying the correlation network tools; in 
 ``Persistence and transitions: dynamical analysis of DBHT'' we describe the analyses we have performed and we discuss the results; in ``Discussions'' we draw the conclusions and discuss future perspectives.

\section*{Correlation-based networks: an overview}
Over the last 15 years correlation-based networks have been used extensively in the Econophysics literature as tools to filter and analyse financial market data
 \citep{mantegna1}, \citep{asset_graphs}, \citep{PMFG2}, \citep{Tumminello05}, \citep{DiMatteo02}, \citep{DiMatteo04}, \citep{DiMatteo05}, \citep{bartolozzi2007multi}.

The seminal work of Mantegna \citep{mantegna1} exploited for the first time a tool from network theory, the Minimum Spanning Tree (MST) (see for instance in \cite{graph_theory}), to analyse and filter from noise the correlation structure of a set of financial assets.
The idea of Mantegna was to look at a correlation matrix as the adjacency matrix of a network and to generate a MST on this network, in order to retain the most significant links/entries. Moreover,
 after mapping the correlation into a suitable metric distance, the MST algorithm provides also a hierarchical classification of the stocks.  
 
In the following years other correlation-based networks have been studied in the literature. 
In \cite{asset_graphs} the authors introduced the dynamic asset graph. Unlike the MST, that filters the correlation matrix according to 
a topological constraint (the tree-like structure of the MST), the dynamic asset graph retains all the links such that the associated correlation (distance) is above (below) a given threshold. In this way it is less affected by 
non-significant, low correlations that are instead often kept by the MST. As a result the dynamic asset graph is more robust against time \citep{asset_graphs}. 
On the other hand, the MST, retaining both high and low correlations,
is more able to uncover global, multi-scale structures of interaction.
Indeed, in financial -- and complex systems in general, several length scales coexist and thresholding at a given value introduces artificially a characteristic size that might hide effects occurring at other scales.


The tree structure exploited in the MST tool is not the only topological constraint that can be used to filter information. In particular if we replace the request of absence of loops with the planarity condition we obtain the 
PMFG \citep{PMFG2}. 
The PMFG can be seen as a generalization of the MST, that is able to retain a higher amount of information \citep{Tumminello05,MatlabPMFG}, having a less strict topology constraint allowing to keep a larger number of links. It can be shown that the hierarchical properties of the MST are preserved in the PMFG. 

We can take this concept a step further and generalize the PMFG to a broader class of networks, by means of the concept of ``genus'' \citep{PMFG2}. 
The genus $g$ of a surface is the largest number of non-intersecting simple closed cuts that can be made on the surface without disconnecting a portion (equal to the number of handles in the surface). 
Requiring that a network be planar, as for the PMFG, is equivalent to requiring that the network can be embedded on a surface with $g = 0$ (ie, no handles,  a topological sphere). 
The natural generalization of the PMFG are therefore networks embedded on surfaces with genus higher than zero. 
The higher the genus, the more handles are in the surface and the more links we can retain from the original correlation matrix. 
More links retained means more information and network complexity, but it means also more noise. 
When $g= \lceil \frac{(N-3)(N-4)}{12} \rceil$ (where $N$ is the number of nodes and $\lceil x \rceil$ is the ceiling function that returns the smallest integer bigger or equal to $x$) the original, fully connected complete graph associated with the correlation matrix can be recovered. 
The concept of embedding on surfaces provides therefore a quantitative way to tune the degree of information filtering by means of a single parameter, $g$, linking correlation-based networks to algebraic geometry  \citep{exploring_genus}.
 
Correlation filtered networks are associated with clustering methods.
Indeed the MST is strictly related \citep{mantegna2} to a hierarchical clustering algorithm, namely the Single Linkage (SL), which can be seen as a network representation of the hierarchy generated by the SL. 
It has been recently shown that a hierarchical clustering can be derived from the PMFG as well \citep{DBHT,MatlabDBHT}. 
The new method is called Directed Bubble Hierarchical Tree (DBHT). 
However the approach is different from the agglomerative one adopted in the Linkage methods: the idea of DBHT is to use the hierarchy hidden in the topology of a PMFG, due to its property of being made of three-cliques \citep{DBHT,bubble}. 
The DBHT hierarchical clustering has been applied to synthetic and biological data in \cite{DBHT} and to financial data in \cite{musmeci_DBHT}, showing that it can outperform many other clustering methods.

\begin{figure}[ht!]
\begin{subfigure}[h]{0.18\textwidth}
   \includegraphics[scale=0.4]{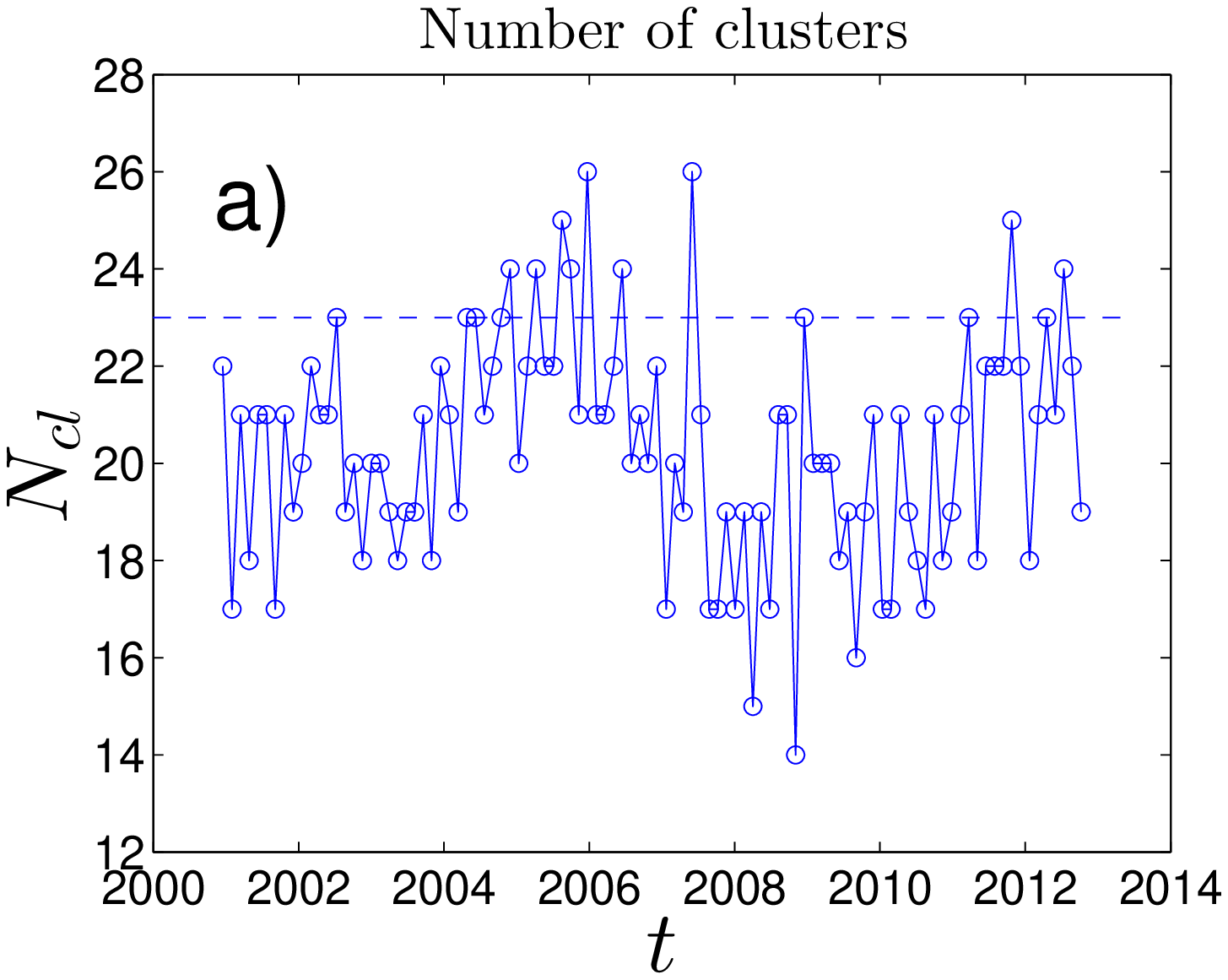}
\end{subfigure}
~~~~~~~~~~~~~~~~~~~~~~~~~~~~~~~~~~~
\begin{subfigure}[h]{0.18\textwidth}
   \includegraphics[scale=0.4]{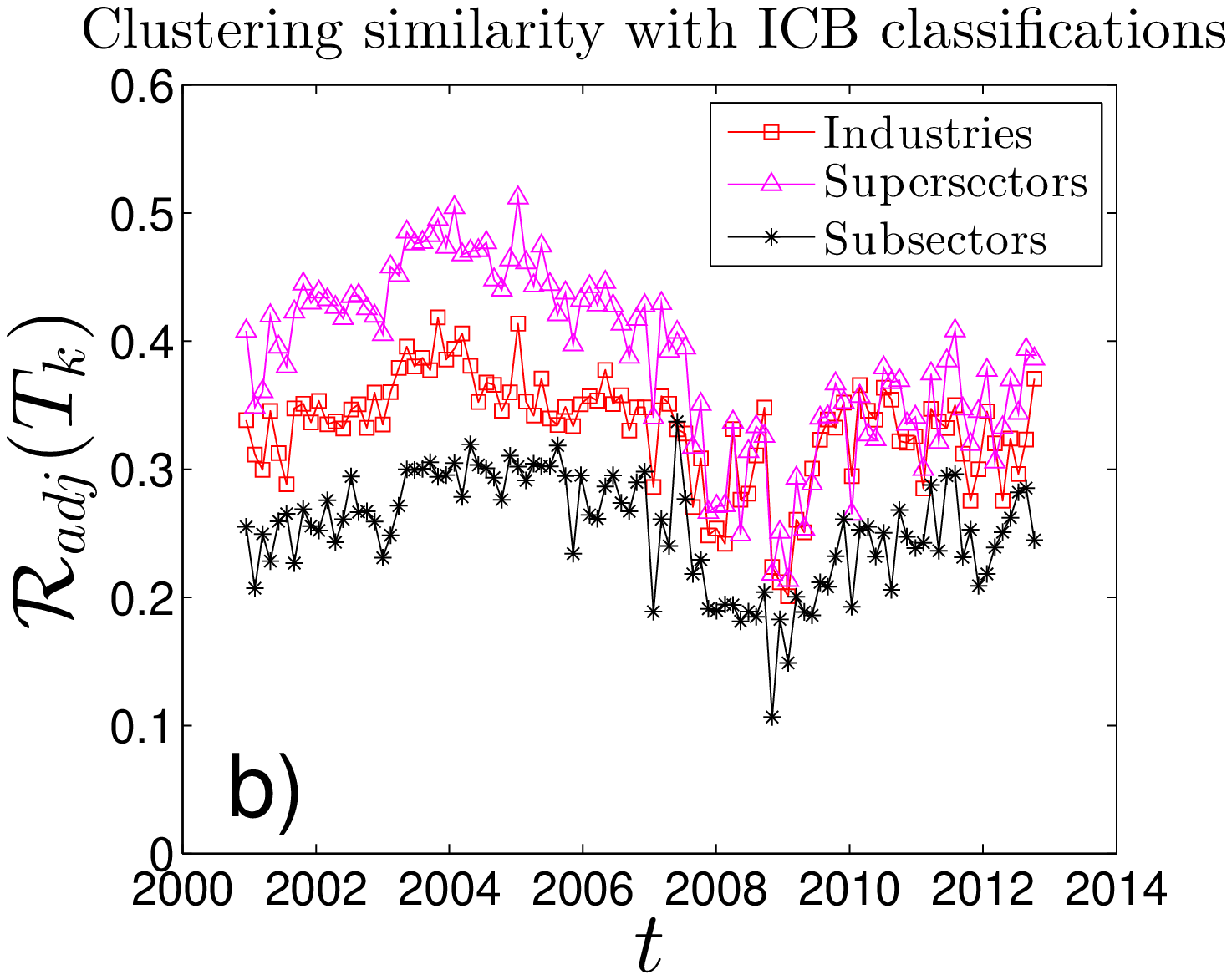}
\end{subfigure}
\\

~~~~~~~~~~~~~~~~~~~
\begin{subfigure}[h]{0.2\textwidth}
   \includegraphics[scale=0.4]{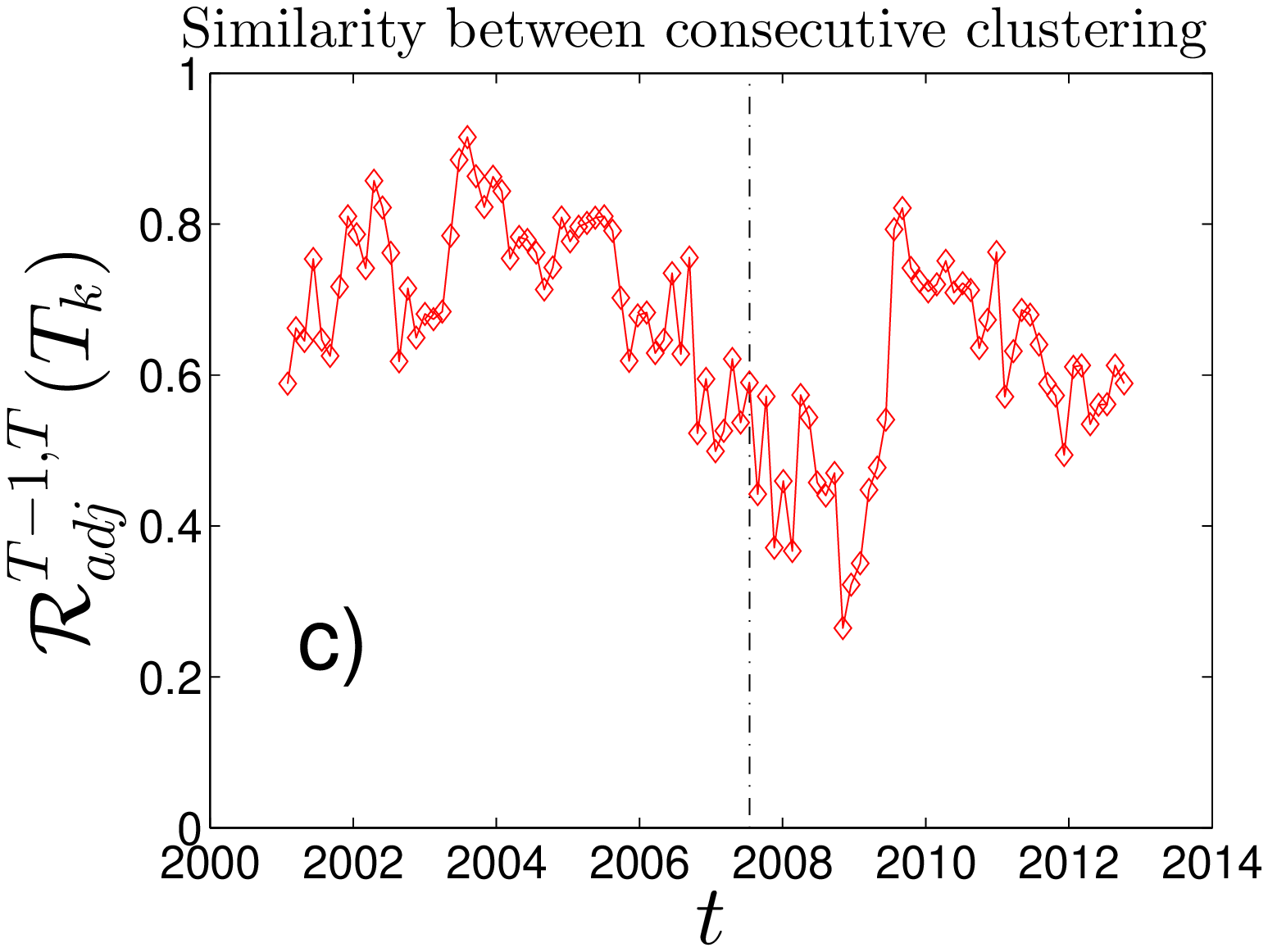}
\end{subfigure}

\caption{\label{fig:num_cl_dbht} {\bf Dynamical evolution of the DBHT clustering}. Each plot refers to 100 moving time windows ($T_k$) of length 1000 trading days and shifted of 30 days.
a) Number of DBHT clusters, $N_{cl}$, with the dashed horizontal line being the $N_{cl}$ value  obtained by taking the entire time window of 4026 trading days (covering years 1997-2012). Overall we can observe a drop in correspondence with the 2007-2008 financial crisis.
b) Amount of economic information retrieved by DBHT clustering in terms of similarity between clustering and ICB partitioning calculated by using the Adjusted Rand Index.  
Again a drop at the outbreak of crisis appears. 
Over the post-crisis years the economic information is less than in the pre-crisis period and differences among different ICB levels are less evident.
c) Persistence of DBHT clustering in time, measured as the Adjusted Rand Index between two adjacent clusterings.
 The financial crisis is characterized by very low levels of persistence.} 
\end{figure} 

\section*{Persistence and transitions: dynamical analysis of DBHT}

We have studied the dynamical evolution of the DBHT clustering on a system of $N = 342$ US stocks during the time period from January 1997 to December 2012.
 To this aim we have selected a set of $n=100$ overlapping time windows $T_k$ with $k=1,...,n$ (each one of length $L=1000$ trading days, with $30$ trading days shift between adjacent time windows)
and computed the distance matrix $D_{ij}(T_k)=\sqrt{2(1-\rho_{ij}(T_k))}$ \citep{mantegna1}, where $\rho_{ij}$ is the following Pearson correlation coefficient: 

  \begin{equation}
  \label{eq:correlation}
  \rho_{ij}(T_k) = \frac{\langle c_i(t) c_j(t) \rangle_{T_k}}{\sqrt{[\langle c^2_i(t) \rangle_{T_k} - \langle c_i(t) \rangle_{T_k}^2][\langle c^2_j(t) \rangle_{T_k} - \langle c_j(t) \rangle_{T_k}^2] }}
 \end{equation}
$\langle ... \rangle_{T_k}$ represents the average over the time window $T_k$ and $c_i(t)$, $c_j(t)$ are the daily log-returns of stocks $i$ and $j$ detrended of the average return. Following \cite{market_mode} we have computed 
$c_i(t)$ for each stock $i$ assuming the following one-factor model for the stock log-return $r_i(t)$:

 \begin{equation}
    r_i(t) = \alpha_i + \beta_i I(t) +c_i(t)
    \label{eq:market_mode}
   \end{equation}
   
where the common market factor $I(t)$ is the market average returns, $I(t)=  \sum_{\gamma=1}^N r_{\gamma}(t)$.
By means of a linear regression we can work out the coefficients $\alpha_i$, $\beta_i$ and finally evaluate $c_i(t)$ as the residual. 
 In agreement with \citep{market_mode} we have verified that correlations on detrended log-returns provide a richer and more robust clustering 
 that can carry information not evident in the original correlation matrix \citep{market_mode}.
We also used a weighted version of the Pearson estimator \citep{exp_smoothing} in order to mitigate (exponentially) excessive sensitiveness to outliers in remote observations.
The DBHT clustering is  calculated on each distance matrix $D(T_k)$.
 
 In Fig. \ref{fig:num_cl_dbht} a) we show 
the number of DBHT clusters obtained for each time window. The number of clusters ranges between 14 and 26. The dashed line is the value (23) correspondent to the clustering obtained using the entire period 1997-2012 
as time window. 
As one can observe, the lowest values are associated with the period around the 2007-2008 financial crisis. 

In order to analyse the amount of economic information expressed by the clustering \citep{mantegna1}, \citep{sect_LSTE} we have measured the Adjusted Rand Index $\mathcal{R}_{adj}$
\citep{adj_rand} between the
DBHT clustering at time window $T_k$ and the community partition generated by the industrial sector classification of stocks. $\mathcal{R}_{adj}$ is an index that measures the similarity between two different partitions on the same
set of objects (stocks in this case) and ranges from $0$ (no similarity) to $1$ (complete identity). We provide a formal definition of this index in Appendix A. $\mathcal{R}_{adj}$ provides therefore a measure of the industrial information contained in the correlation based clustering.
We use the Industrial Classification Benchmark (ICB); which is a categorization that divides the stocks in four hierarchical levels, namely in 114 subsectors, 41 sectors, 19 different supersectors, that, in turn, are gathered in 10 different Industries.
In order to take into account all these levels we have measured $\mathcal{R}_{adj}(T_k)$ between each of these hierarchical levels and the DBHT clustering.
In Fig. \ref{fig:num_cl_dbht} b) we plot the evolution in time of $\mathcal{R}_{adj}(T_k)$  between DBHT clusters and 
ICB industries, supersectors and subsectors (for sake of simplicity we do not plot the sectors data that are very close to supersectors values). One can see how the ICB information shows a remarkable drop during the 2007-08 
financial crisis, to be partially recovered from 2010 onwards. Interestingly before the crisis the industry, supersector and subsector lines were clearly distinct (with ICB supersectors showing the highest similarity with DBHT,
 followed by industries and subsectors) whereas in the crisis and post-crisis periods they display much closer values. Therefore from the crisis onwards the correlation clustering is no longer able to distinguish between different 
 levels of ICB: this might indicate that this industrial classification is becoming a less reliable benchmark to diversify risk. 
 These results are confirmed with other industrial partitionings, including the Yahoo classification.  
 
 The Adjusted Rand Index can also be used as a tool for analyzing the persistence of DBHT clustering by measuring the index between two clusterings at two adjacent time windows (we denote  $\mathcal{R}_{adj}^{T-1,T}(T_k)$ such a quantity). 
 This gives a measure of local persistence: a drop in the index value indicates decreasing similarity between adjacent clusterings, and therefore less persistence.
 In Fig. \ref{fig:num_cl_dbht} c) we plot  
  $\mathcal{R}_{adj}^{T-1,T}(T_k)$ against time. We can observe that the clustering persistence changes remarkably over time, dropping in particular with the outbreak of financial crisis and recovering in 2010. 
  It is worth pointing out that the drop during the crisis seems to start some months earlier than the actual outbreak of it (August 2007, dashed vertical line): this could highlight a possible use of clustering persistence 
  as tool to forecast systemic risk. 
  The time period 2010-2012 shows however again a steady decreasing trend. Interestingly the pattern of persistence appears to be related to the similarity between clustering and ICB, with periods of higher persistence characterized 
  by higher amount of economic information. 
  
  However the drawback of $\mathcal{R}_{adj}^{T-1,T}(T_k)$ as a measure of persistence is that at any time it only provides information on the persistence with respect the previous, adjacent time window. It tells nothing about
  long-term robustness of each clustering. To investigate this aspect we discuss in the next section a set of analyses that evaluate the persistence of each clustering at each time providing therefore a more complete picture.

 \subsection*{A map of structural changes}

 \begin{figure}[ht!]
 \begin{center}
  \includegraphics[scale=0.4]{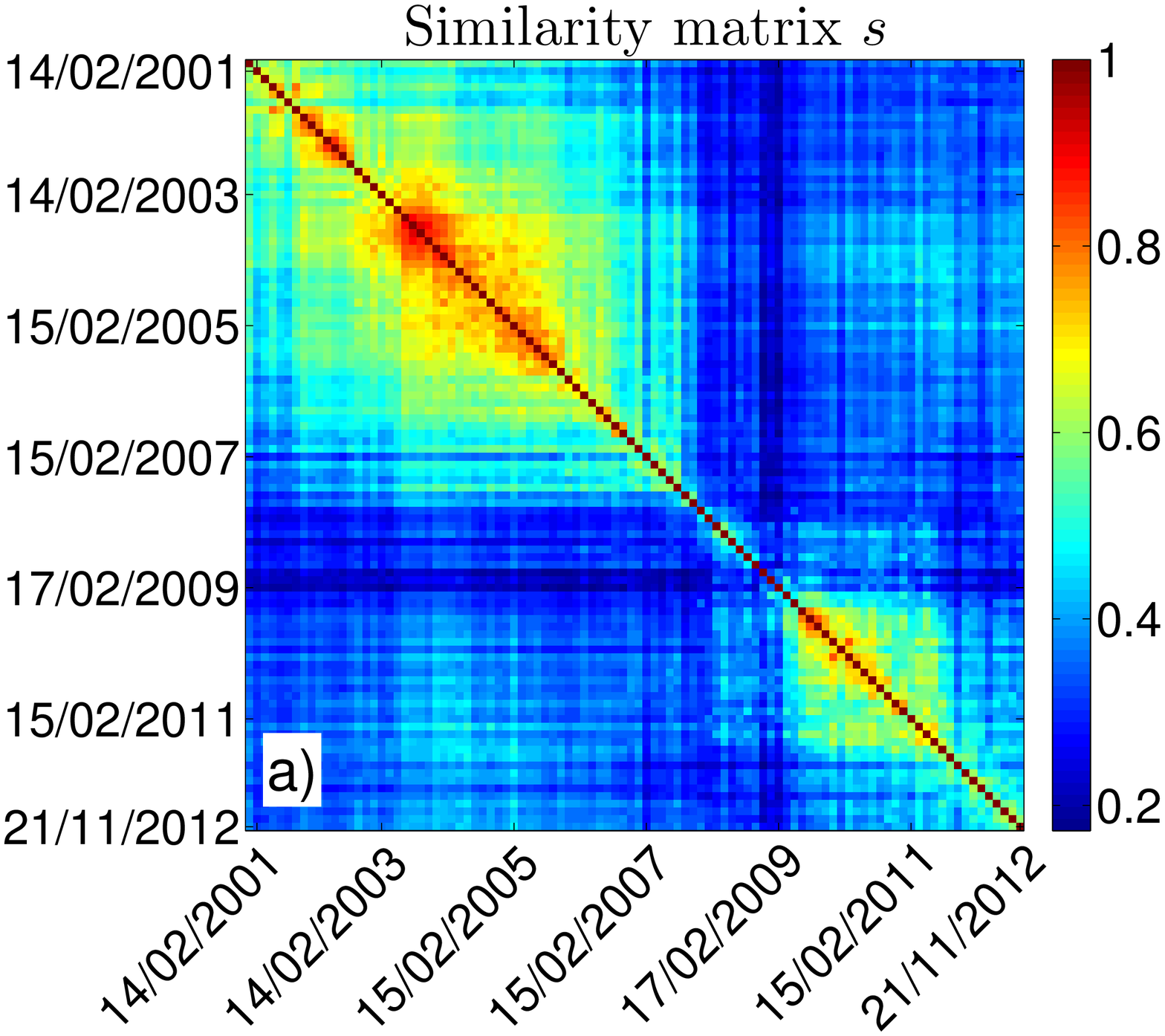}\\
  \vskip-1.5cm
   \includegraphics[scale=0.25]{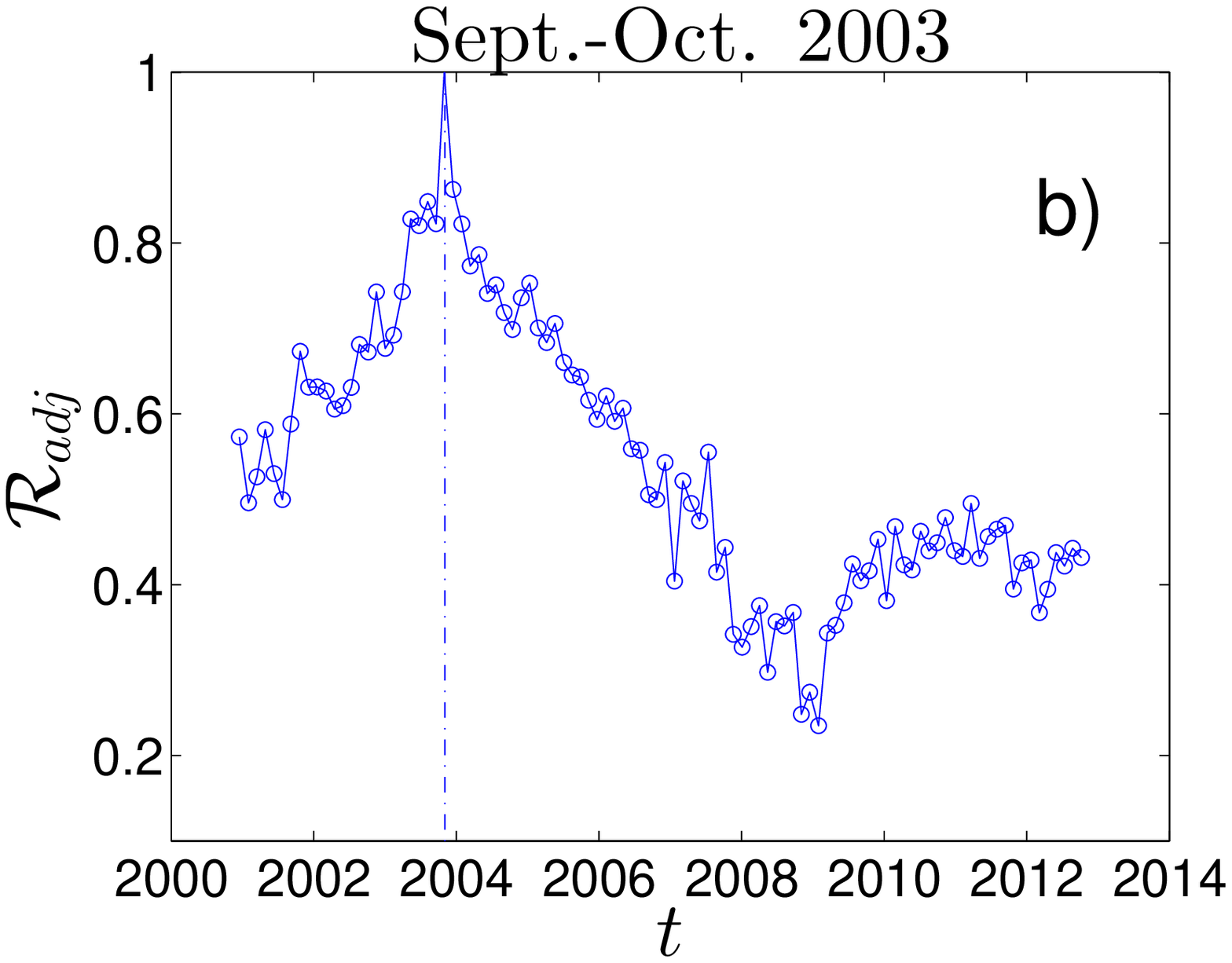}   
   \includegraphics[scale=0.25]{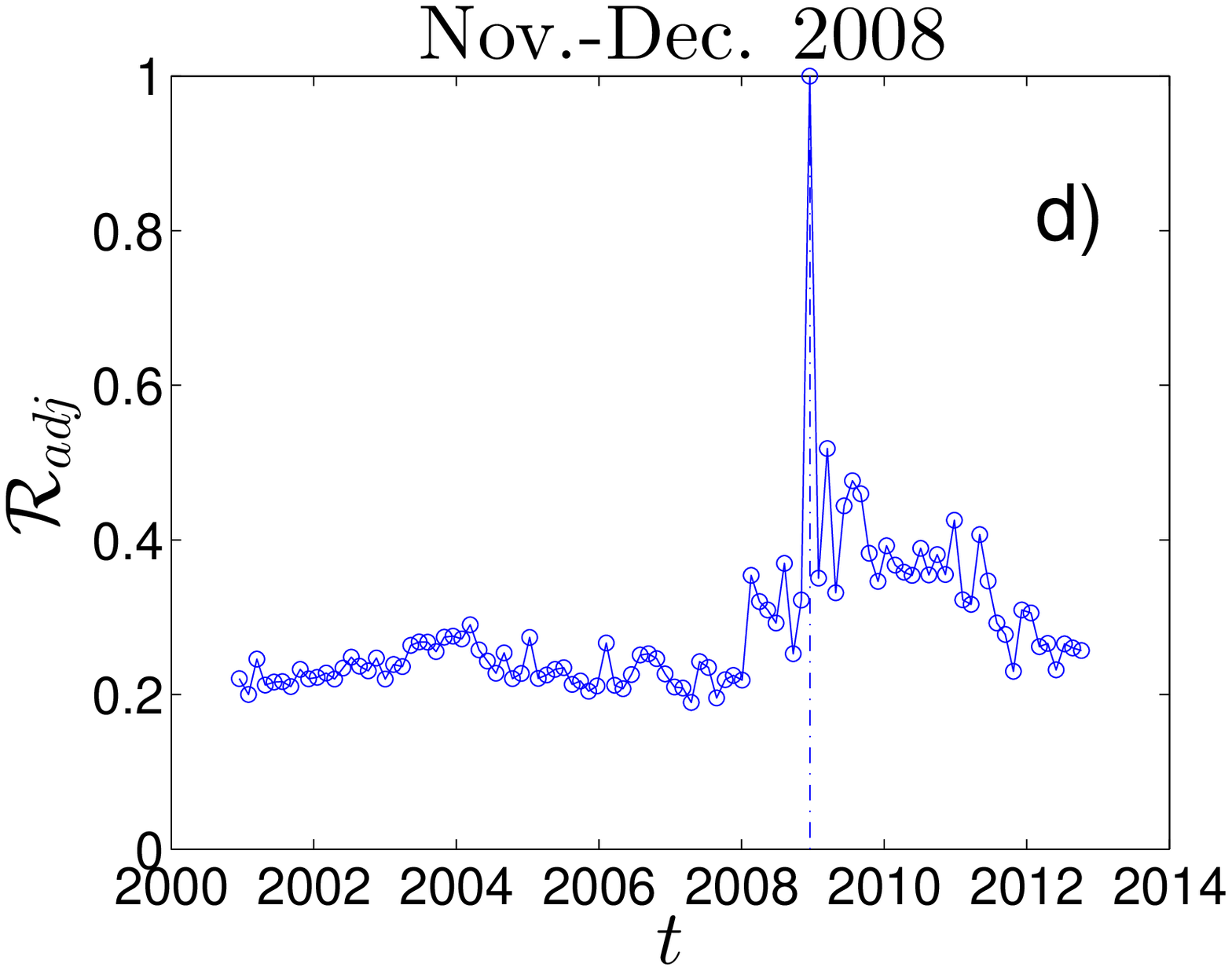}\\
   \includegraphics[scale=0.25]{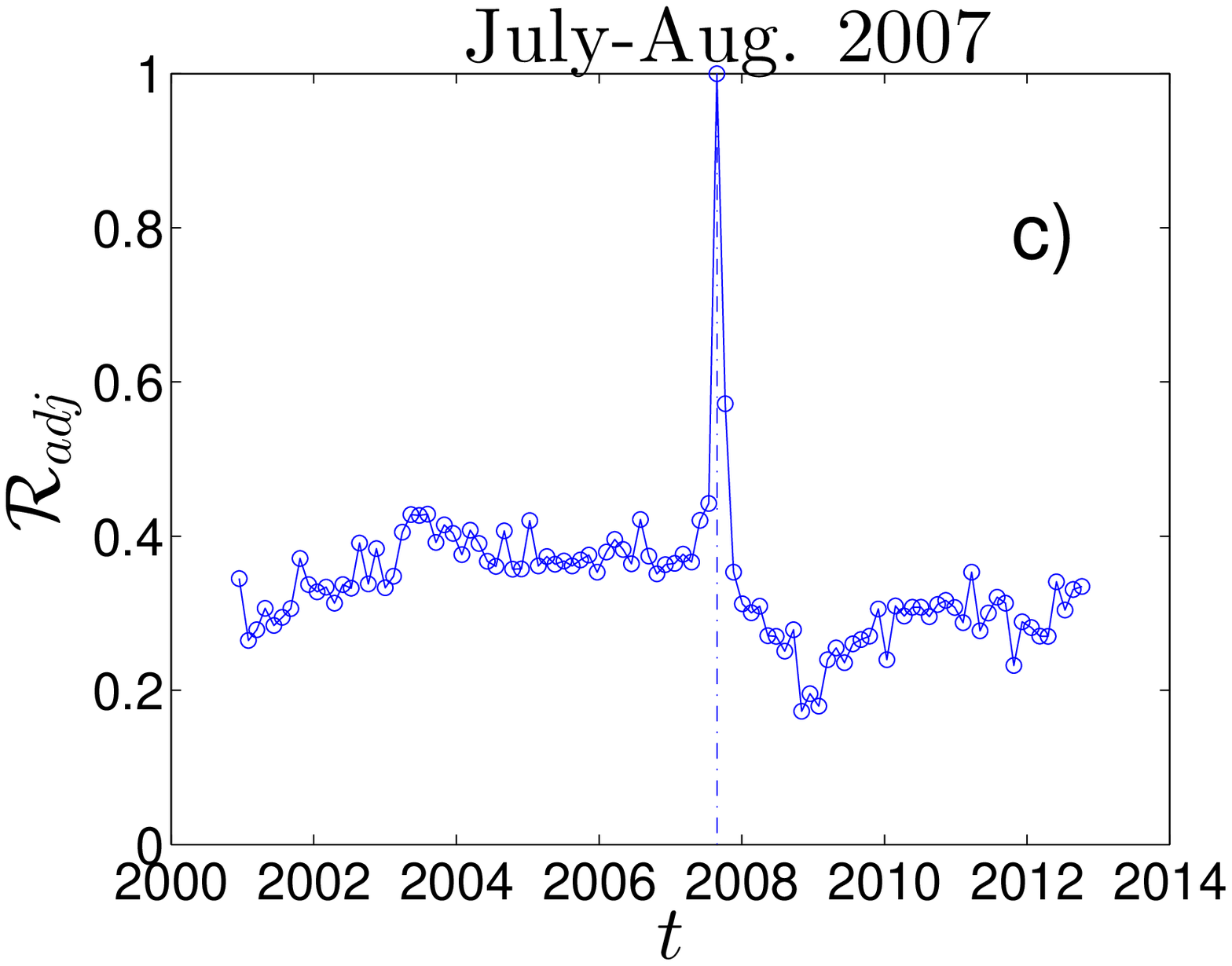}
\includegraphics[scale=0.25]{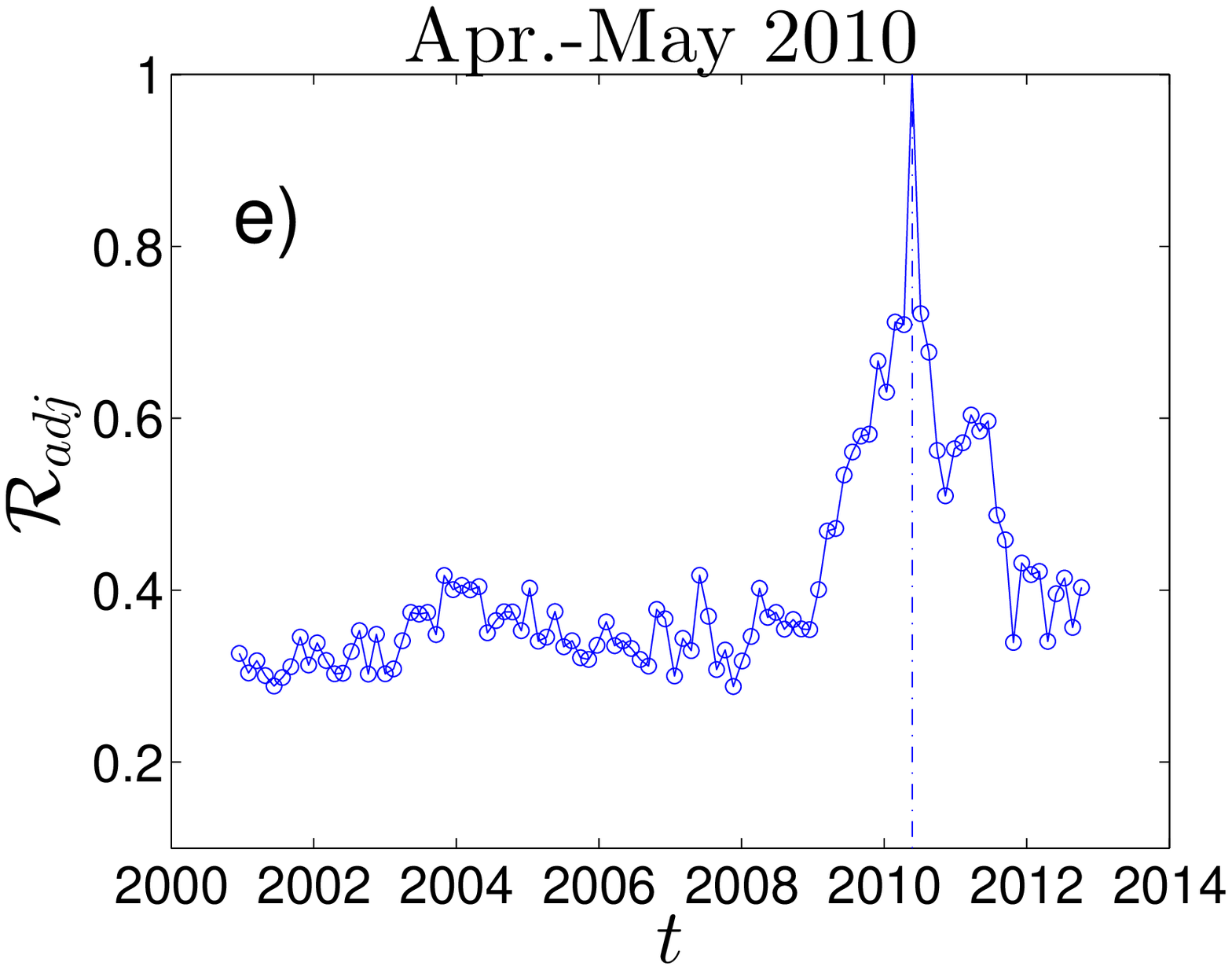}
 \vspace{-.5cm}
\end{center}
%
%

\caption{\label{fig:market_states1} {\bf Persistence analysis based on clustering}. a) Similarity matrix $s$ showing the temporal evolution of the correlation-based DBHT clustering. 
Each entry $s(T_a, T_b)$ is the Adjusted Rand Index between  
clustering $X_a$ and $X_b$ at time window $T_a$ and $T_b$ respectively (Eq. \ref{eq:market_states2}): higher values indicate higher similarity. The matrix displays two main blocks of high intra-similarity, one 
pre-crisis and the other one post-crisis. The years 2007-2008 fall between these two blocks and display very low similarity with any other time window, revealing an extremely changeable structure. Figures b)-e) show the patterns 
of similarity for four sample time windows (i.e. four sample rows of the similarity matrix): during the crisis the decay of similarity becomes much faster than in the pre and post-crisis periods.} 
\end{figure}

%
%

 \begin{figure}[ht!]
 \begin{center}
  \includegraphics[scale=0.4]{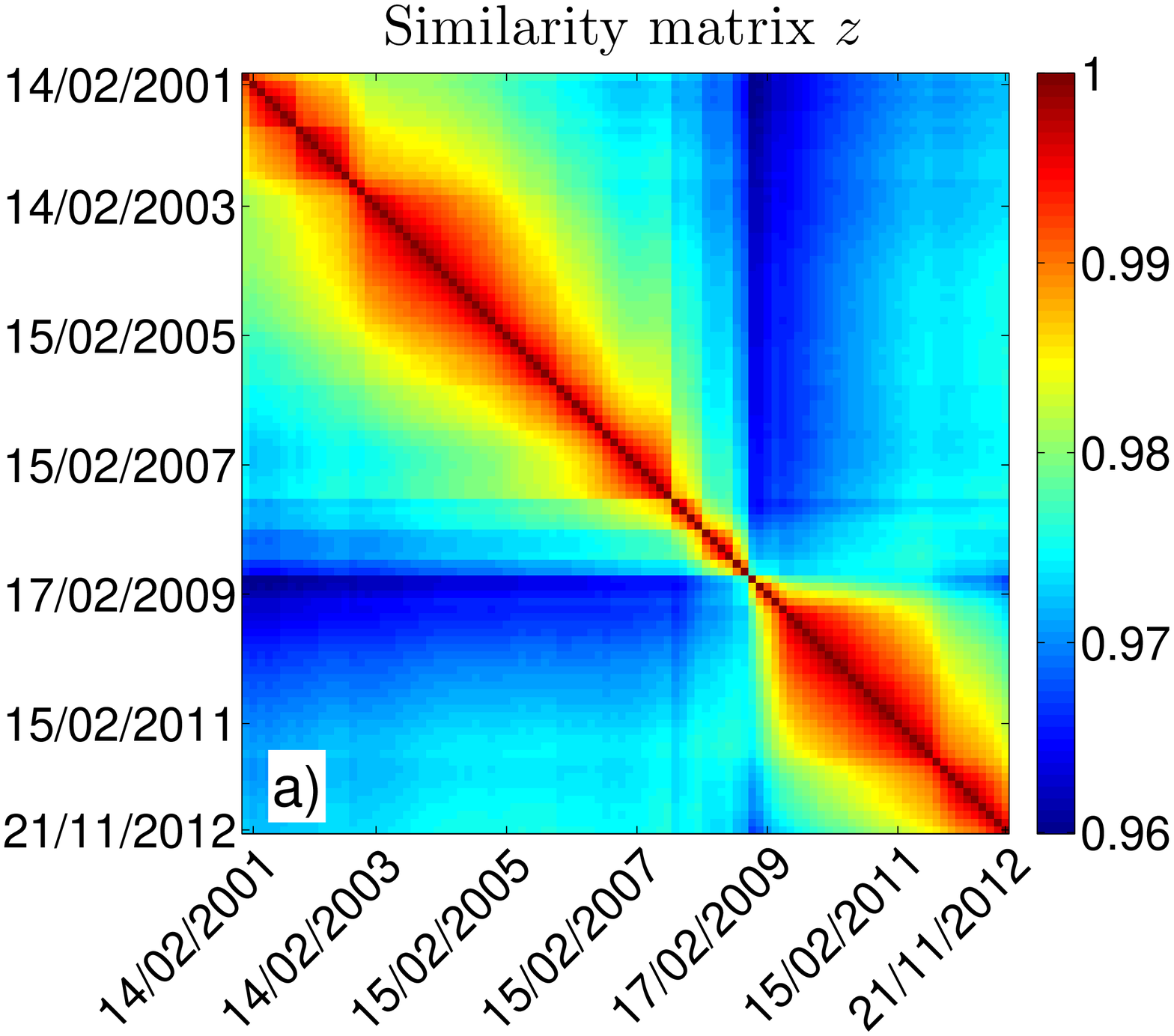} \\
  \vskip-1.5cm
  \includegraphics[scale=0.25]{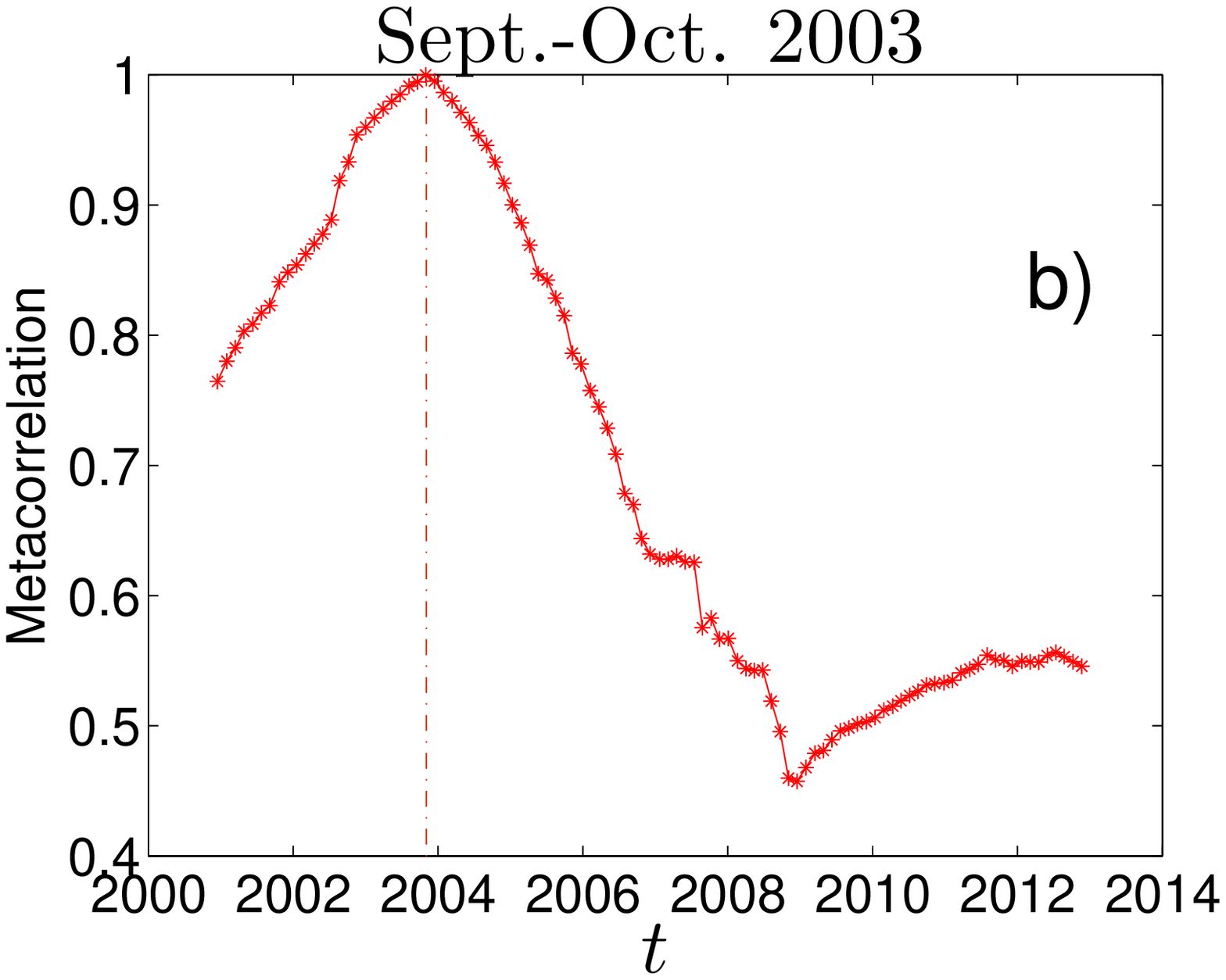}   
  \includegraphics[scale=0.25]{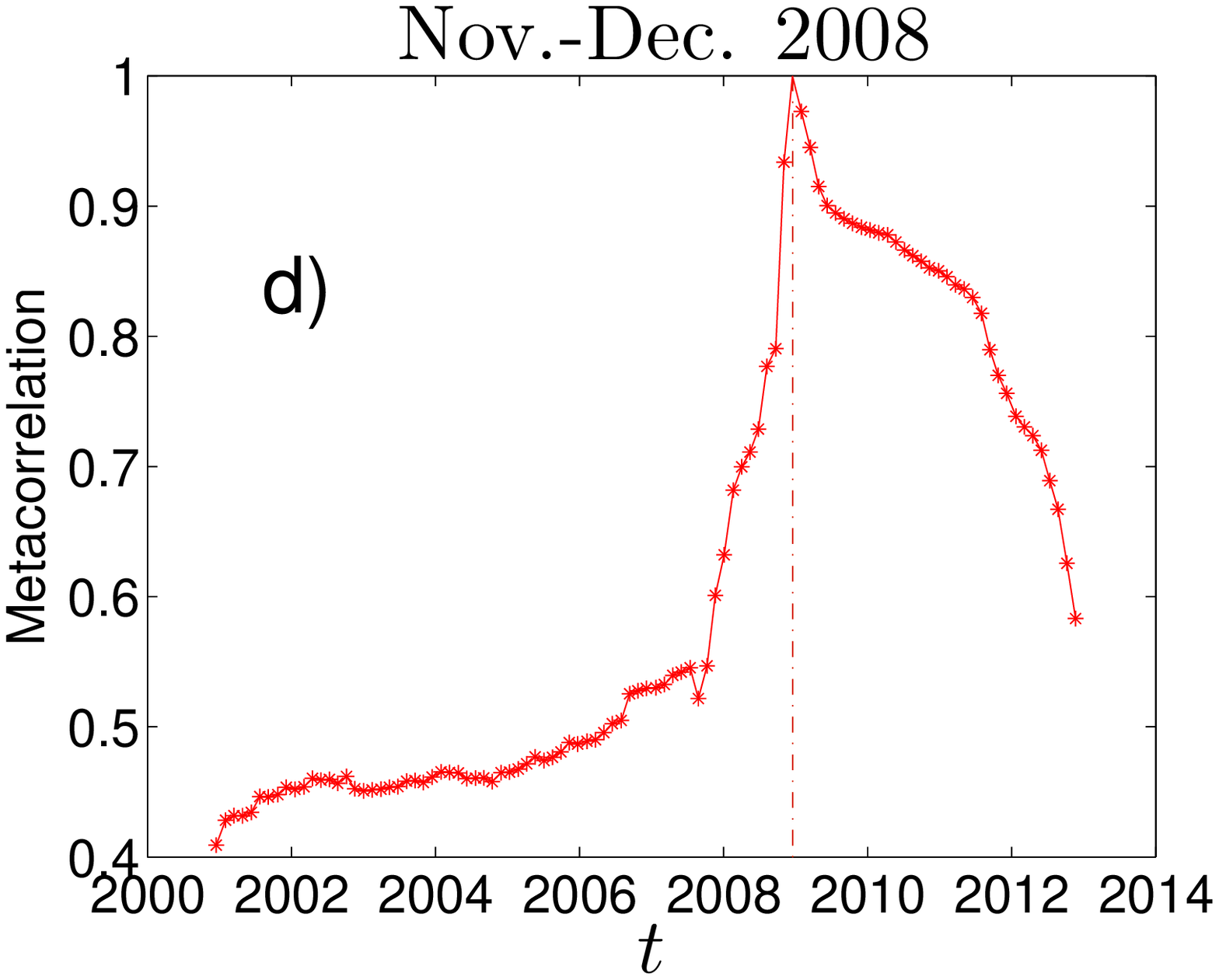} \\
  \includegraphics[scale=0.25]{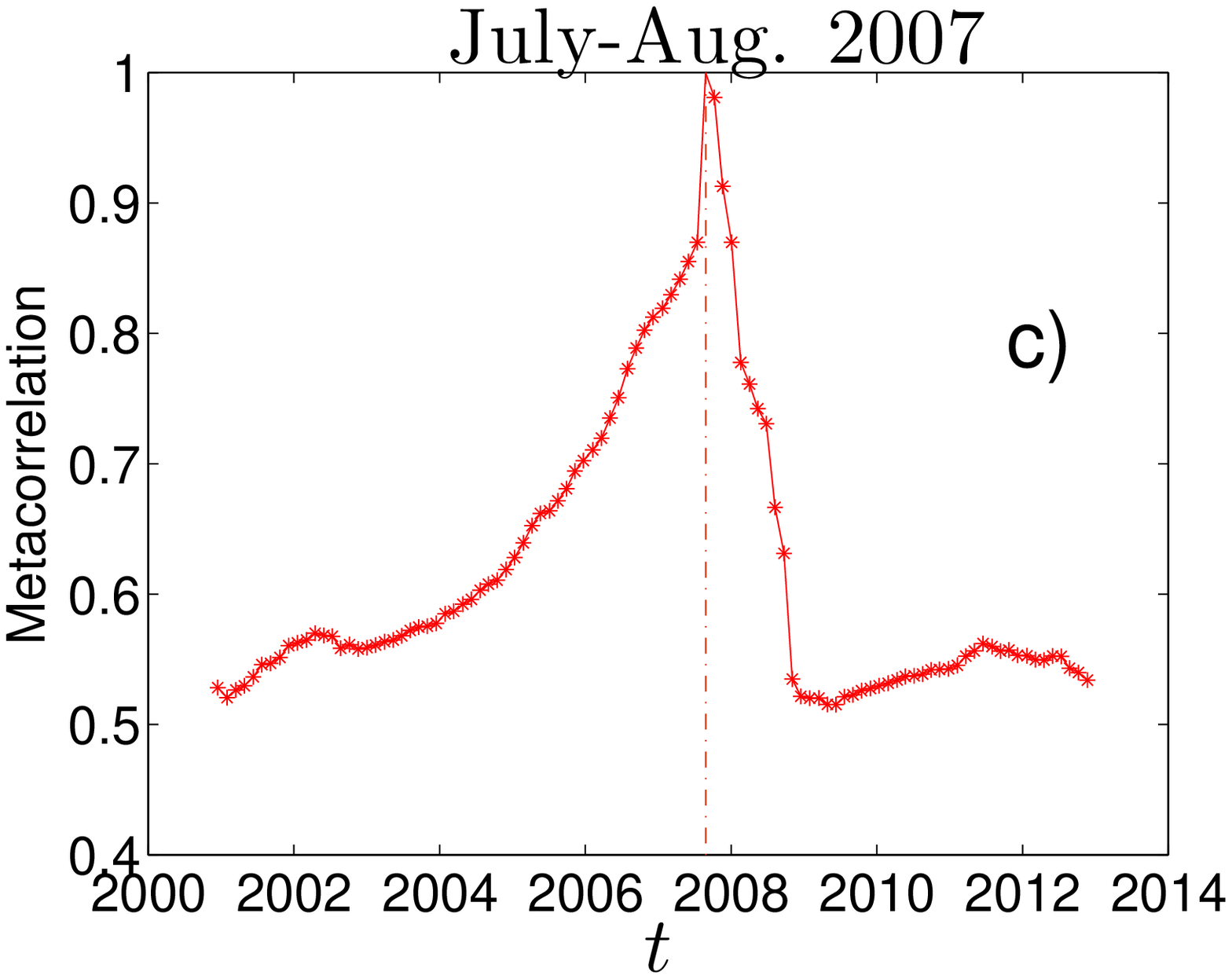}
  \includegraphics[scale=0.25]{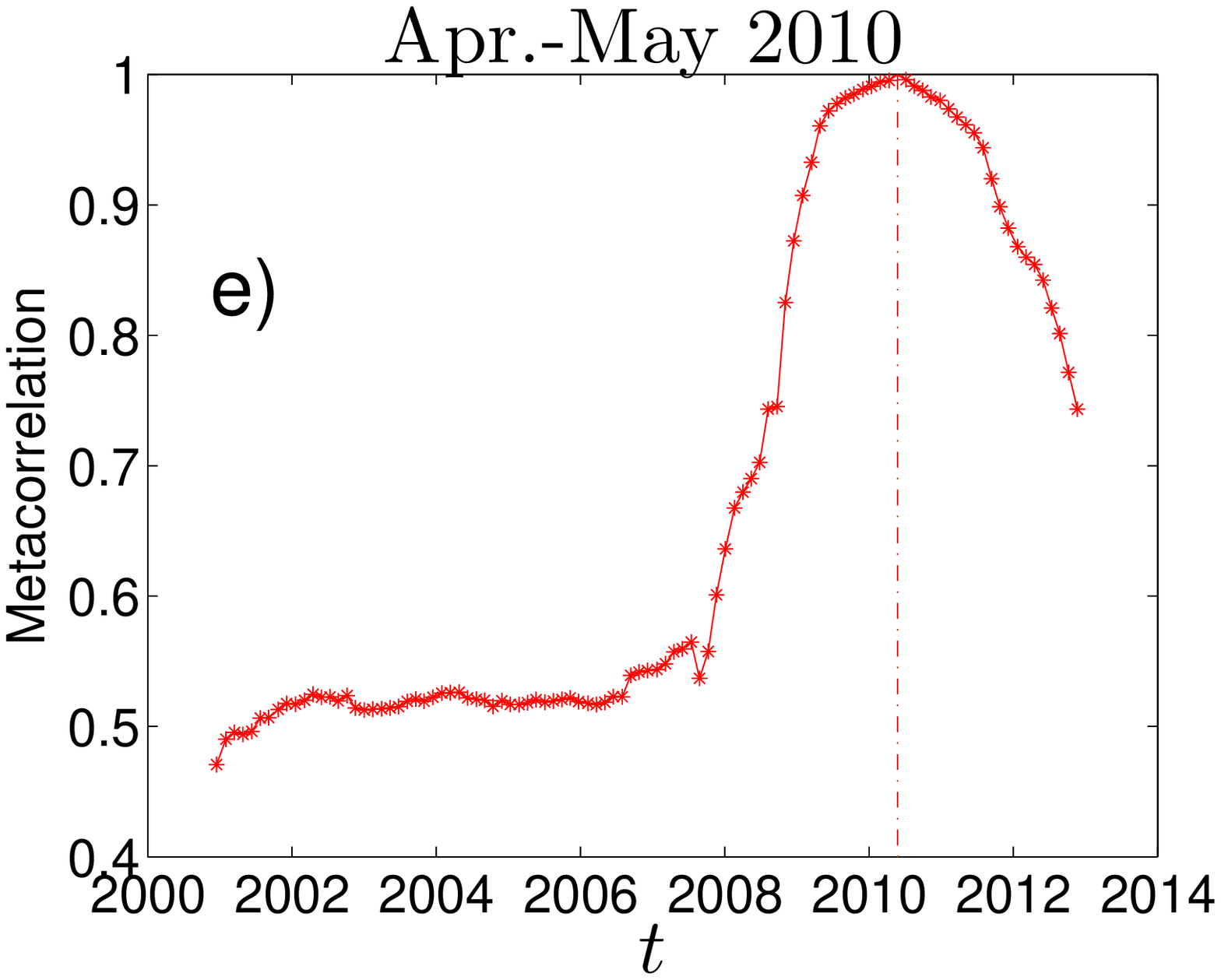}
  \end{center}
%
%

\caption{\label{fig:market_states2} {\bf Persistence analysis based on metacorrelation}. a) Similarity matrix $z$ showing the temporal evolution of correlation matrices. Each entry $z(T_a, T_b)$ is calculated as correlation
among correlation matrices 
at time windows $T_a$ and $T_b$ (Eq. \ref{eq:market_states3}): higher values indicate higher similarity.
Figures b)-e) show the patterns of similarity for four sample time windows: the decay during the crisis years is much less steep than for the corresponding plot 
in Fig. \ref{fig:market_states1}.} 
\end{figure} 

To investigate the long-term persistence of each clustering we have calculated for each time window the Adjusted Rand Index between the correspondent clustering and the clustering at any other time: the result is  summarized in the (symmetric) 
matrix $s$: 

\begin{equation}
  s(T_a,T_b) = \mathcal{R}_{adj}(X_a, X_b)
     \label{eq:market_states2}
 \end{equation}

 where $X_a$ and $X_b$ are the DBHT clusterings at time windows $T_a$ and $T_b$ respectively. The matrix $s$ for our dataset is shown in Fig. \ref{fig:market_states1} a).
 We observe two main blocks, the first pre-crisis and the other post-crisis, within which one find high similarity among clusterings. The two blocks show very low mutual similarity (upper right corner/lower left corner of the matrix).
 The first block begins losing its compactness in 2007, and the second one quite quickly at the beginning of 2009: between these two times the outbreak of financial crisis displays a series of extremely changeable clusterings, that 
 do not show similarity with any other time window.
 
 To better highlight these changes of regime we plot in Figs. \ref{fig:market_states1} b) - e) four time rows from matrix $s$, taken as examples of persistence behaviour during the pre-crisis 
 (September-October 2003, b)), crisis (July-August 2007, the outbreak of crisis, and November-December 2008, the aftermath of Lehman Brothers default, c) and d)) and post-crisis period (April-May 2010).
 The vertical dashed lines show the end position of the time window whose clustering that is taken as reference: each point in the plot 
 is the Adjusted Rand Index between that clustering and all the other clusterings at each other time window, both in the past and the future. In the pre-crisis period b) the similarity displays a quite slow decay both 
 forward and backward in time: the original clustering has still $60\%$ of similarity with the 17th time window forward/backward in time. The decreasing trend is however well evident and becomes steeper during the crisis. 
 Taking time windows during the financial crisis, c) and d), the pattern changes drastically: the similarity drops by $70$-$80\%$ in few months both backward and forward in time. The two stages of crisis reveal also some differences:
 while in the early crisis period c) the similarity with pre-crisis clusterings is higher than with the post-crisis ones, in the post Lehman Brothers period d) the situation is reversed. Finally, the post-crisis period e) shows a partially recovered persistence, although not at the same levels of the 2003 pattern.
 
 One could wonder whether these structural changes highlighted by the clustering analyses can be detected directly by studying the original, unfiltered correlation matrices. 
 To check this we introduce an alternative measure of similarity among different time windows that does not make any use of clustering, namely the correlation calculated between pairs of correlation matrices (metacorrelation). 
 This measure is:
 
 \begin{equation}
  z(T_a,T_b) = \frac{\langle \rho_{ij}(T_a) \rho_{ij}(T_b) \rangle_{ij}}{\sqrt{[\langle \rho^2_{ij}(T_a) \rangle_{ij} - \langle \rho_{ij}(T_a) \rangle_{ij}^2][\langle \rho^2_{ij}(T_b) \rangle_{ij} - \langle \rho_{ij}(T_b) \rangle_{ij}^2] }}
 \label{eq:market_states3}
 \end{equation}

 where $\rho_{ij}(T_a)$ is the correlation between stocks $i$ and $j$ at time window $T_a$ and $\langle ... \rangle_{ij}$ is the average over all couples of stocks $i,j$. In \cite{market_states} an alternative measure has been
 introduced to identify the possible states of a financial market. 
 In Fig. \ref{fig:market_states2} we report
 the matrix $z(T_a, T_b)$ and four representative time rows, corresponding to the same four time windows chosen in Fig. \ref{fig:market_states1}. 
 We can observe that metacorrelation is indeed able to identify the two pre-crisis and post-crisis time blocks, but shows also a smaller, intermediate block during the 2007-2008 crisis with a relative high intra-similarity. 
 This is different from what we have observed in the clustering based matrix $s$, where the time windows during the crisis are quite dissimilar even from each other. Moreover the pre-crisis and post-crisis blocks in $z$ display higher 
 intra-similarity than $s$, especially over the post-crisis years. All these differences can be appreciated looking at the four $z$ time rows in Figs. \ref{fig:market_states2} b)-e): even if in the crisis time windows c) and d)
 a faster decay of similarity can be observed, the decay is much less steep than the corresponding clustering plot (Figs. \ref{fig:market_states1} c) and d)). Moreover the post-crisis window e) recovers completely the high pre-crisis
 level of persistence, unlike the clustering case in Fig. \ref{fig:market_states1} e).  
  
Therefore it seems that metacorrelation and clustering analysis depict slightly different dynamics of market correlation structure. In particular the clustering based matrix $s$ reveals higher non-stationarity during the crisis 
and the post-crisis period.

\subsection*{Clusters composition evolution}

 \begin{figure}[ht!]
 \begin{center}
   \includegraphics[width=0.49\textwidth]{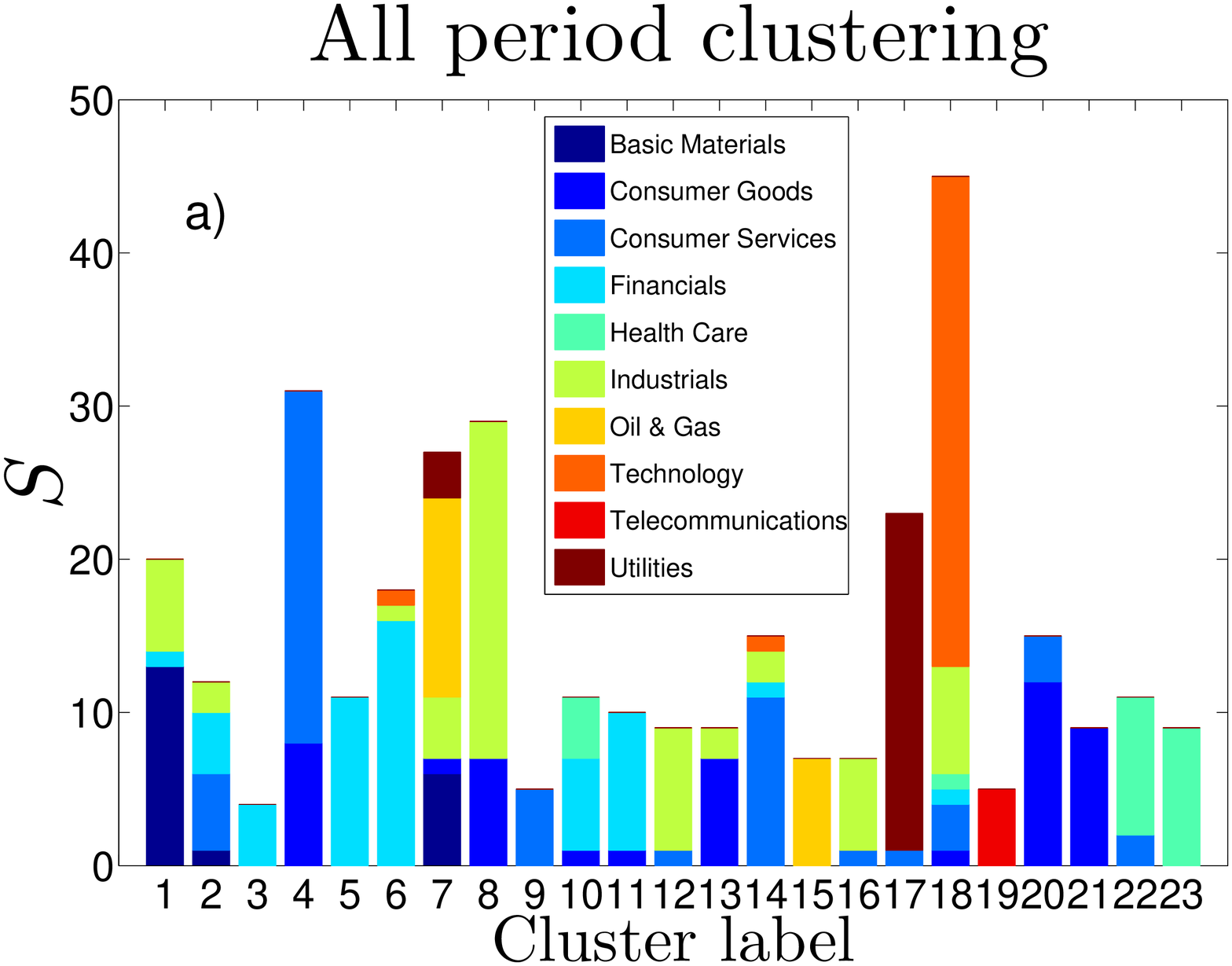}
   \includegraphics[width=0.49\textwidth]{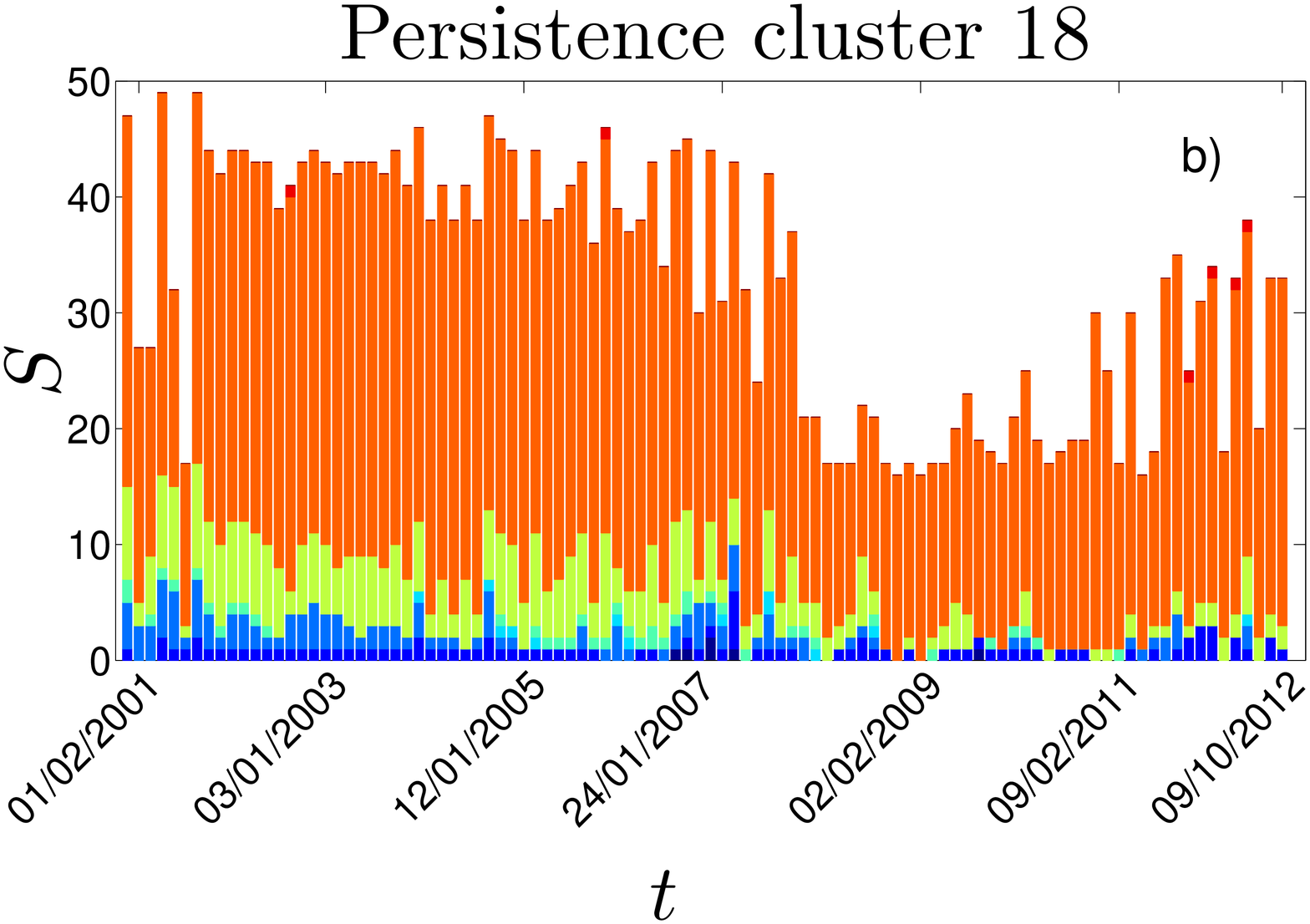}
   \includegraphics[width=0.49\textwidth]{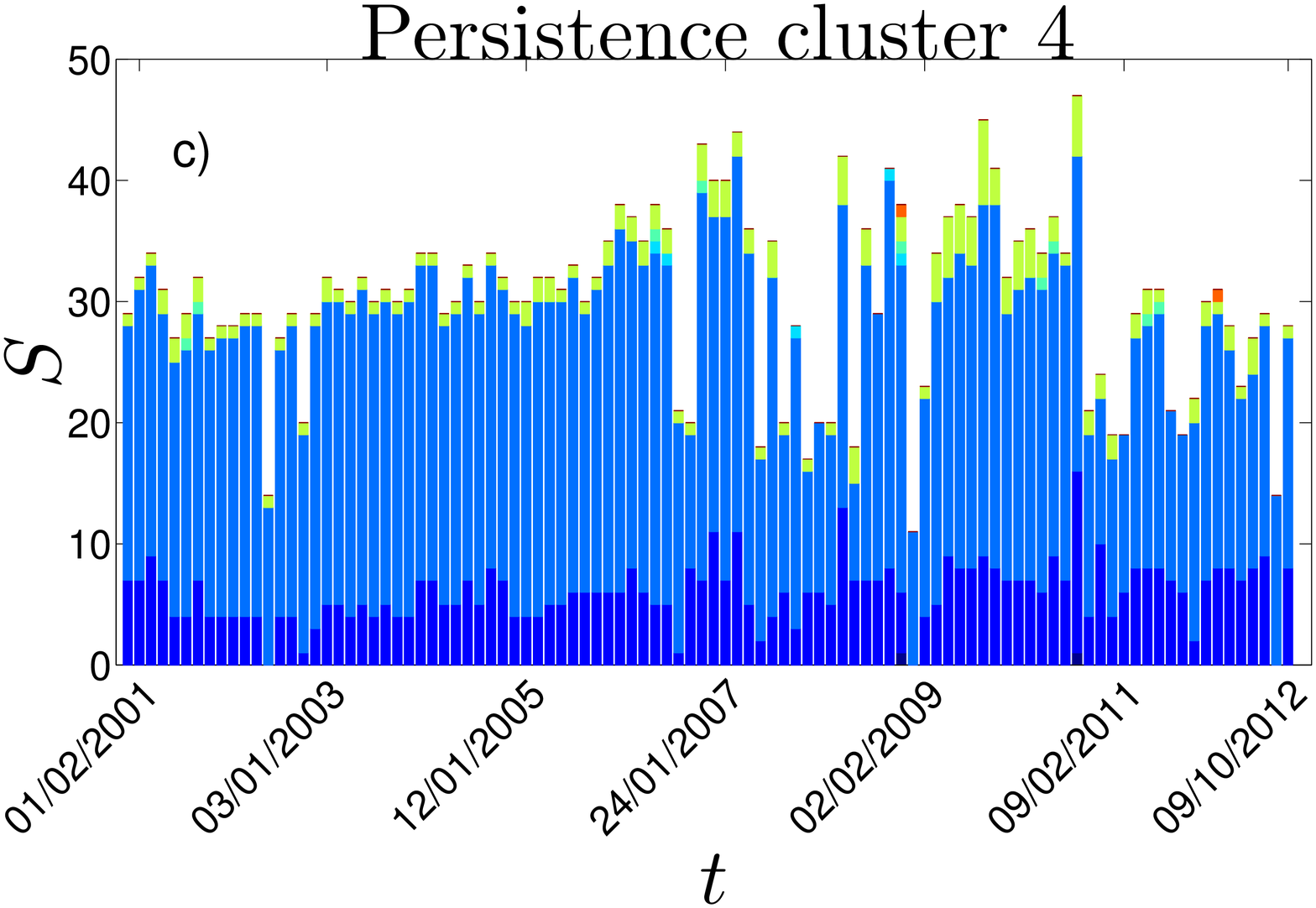}
   \includegraphics[width=0.49\textwidth]{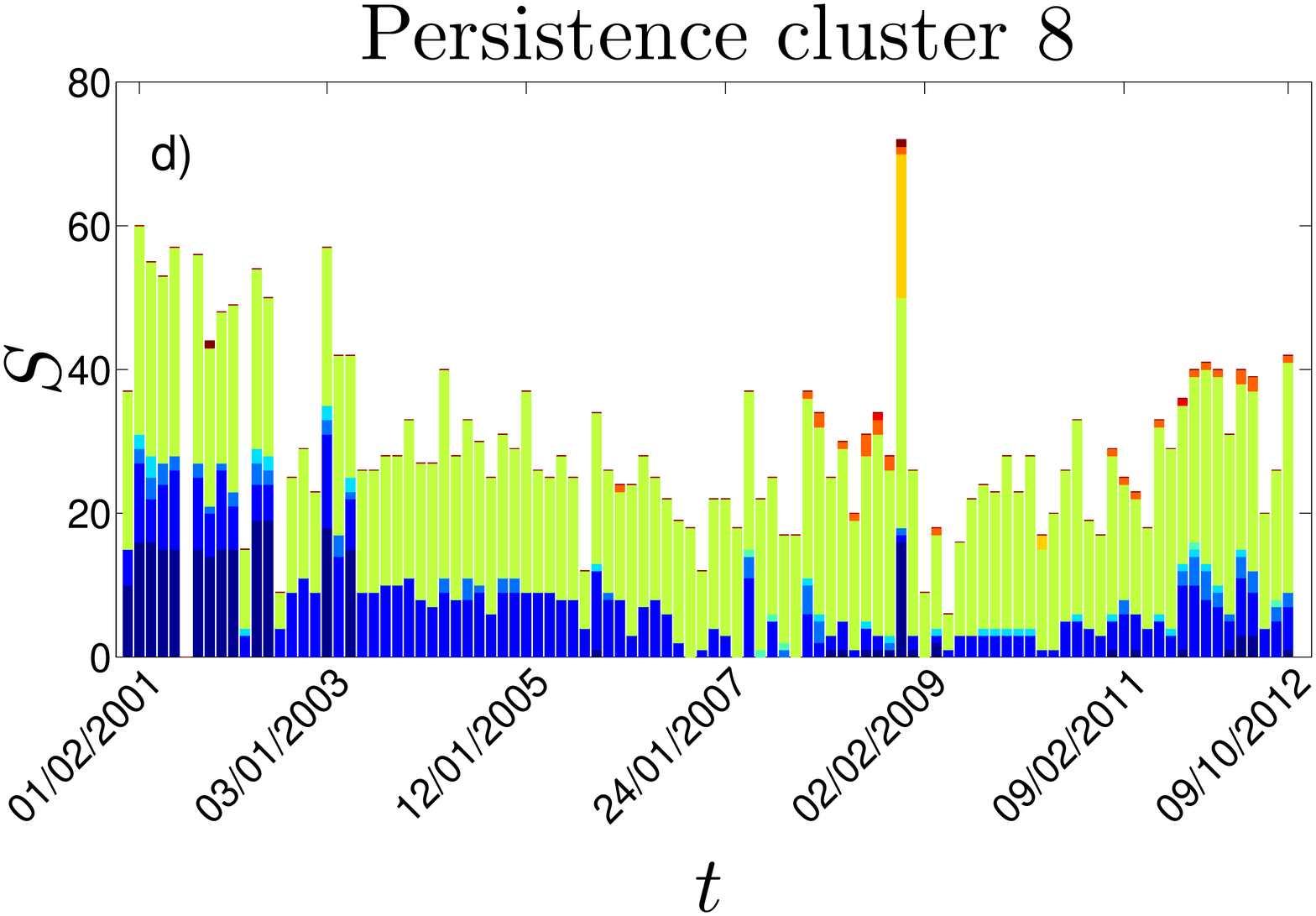}
   \includegraphics[width=0.49\textwidth]{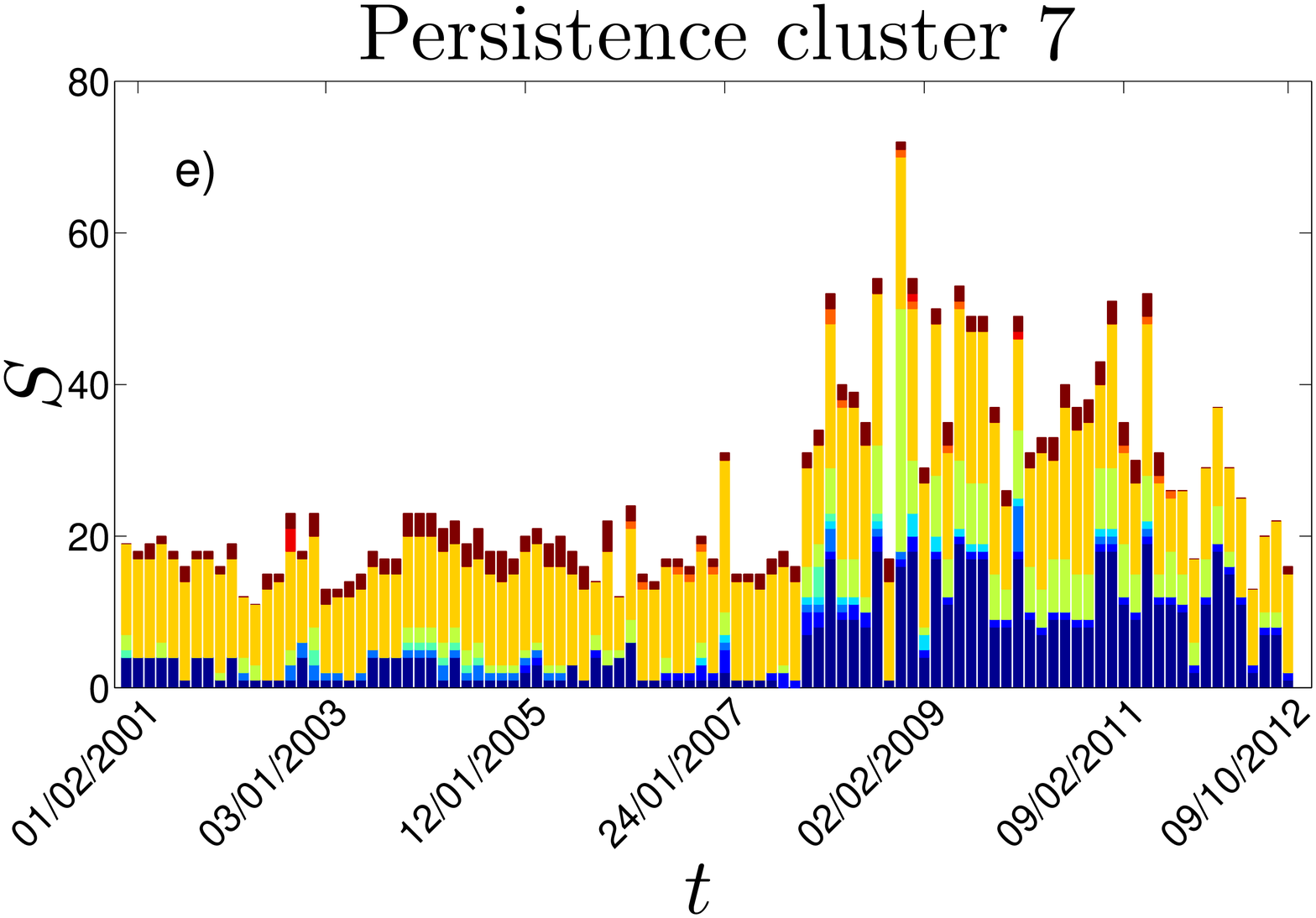}
   \includegraphics[width=0.49\textwidth]{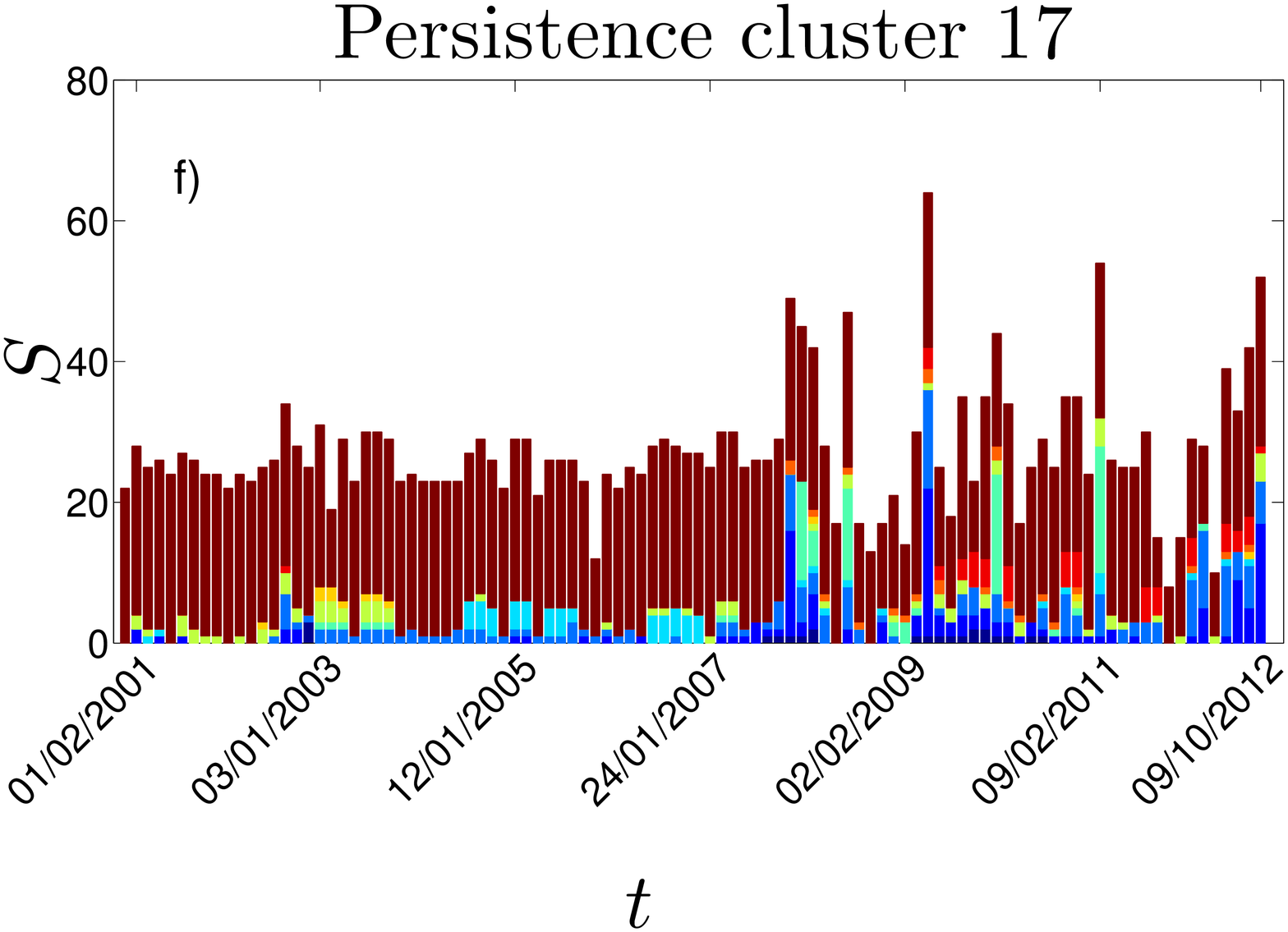}
 \end{center}
%
%
\caption{\label{fig:cl_composition} {\bf Clusters dynamical composition (part 1)}. a) Clusters composition of DBHT clusters obtained by calculating detrended log-returns on the entire time window 1997-2012. On the y-axis the number of 
stocks in each cluster is shown, with different colours for different ICB industries. b) For the cluster number 18 in a) we have detected at each time window the correspondent ``similar'' (according to the hypergeometric test)
cluster and we have plotted the composition in time. Size equal to zero corresponds to no ``similar'' cluster found. When more than one ``similar'' cluster is found only data of the largest cluster is plotted. c)-f): same plots 
as in b), for clusters 4, 8, 7 and 17 respectively.} 
\end{figure} 

 \begin{figure}[ht!]
    \includegraphics[width=0.49\textwidth]{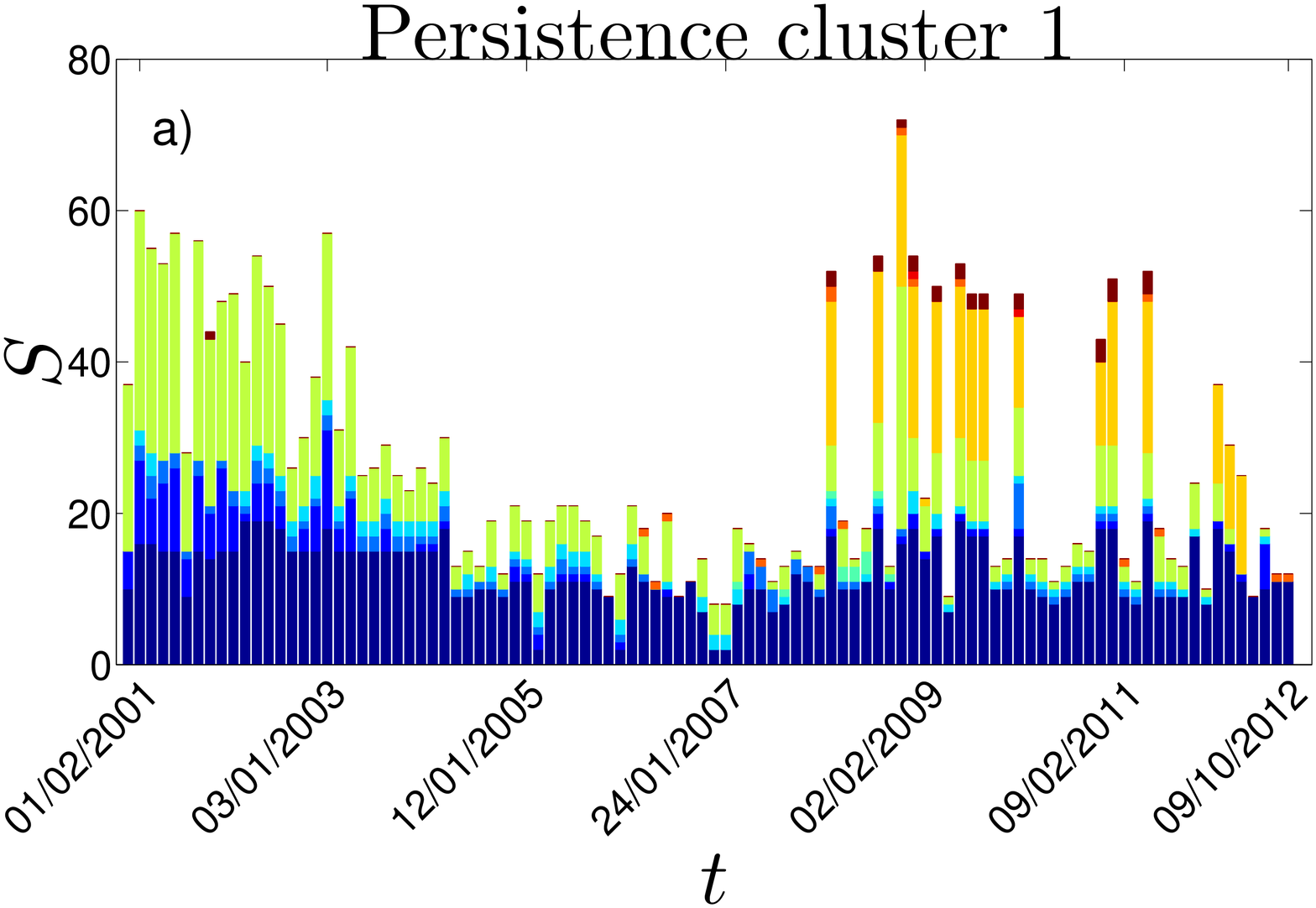}
   \includegraphics[width=0.49\textwidth]{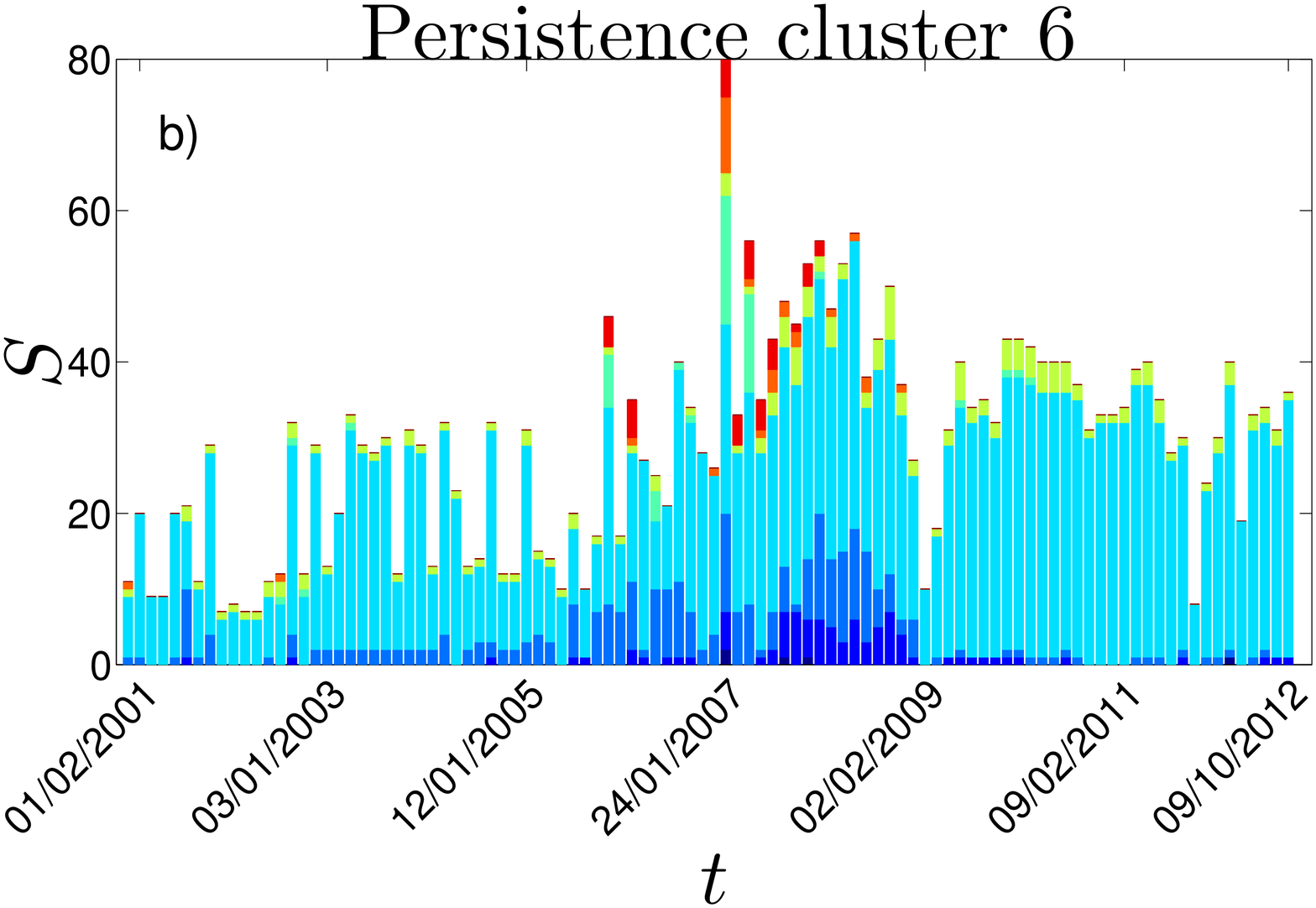}
   \includegraphics[width=0.49\textwidth]{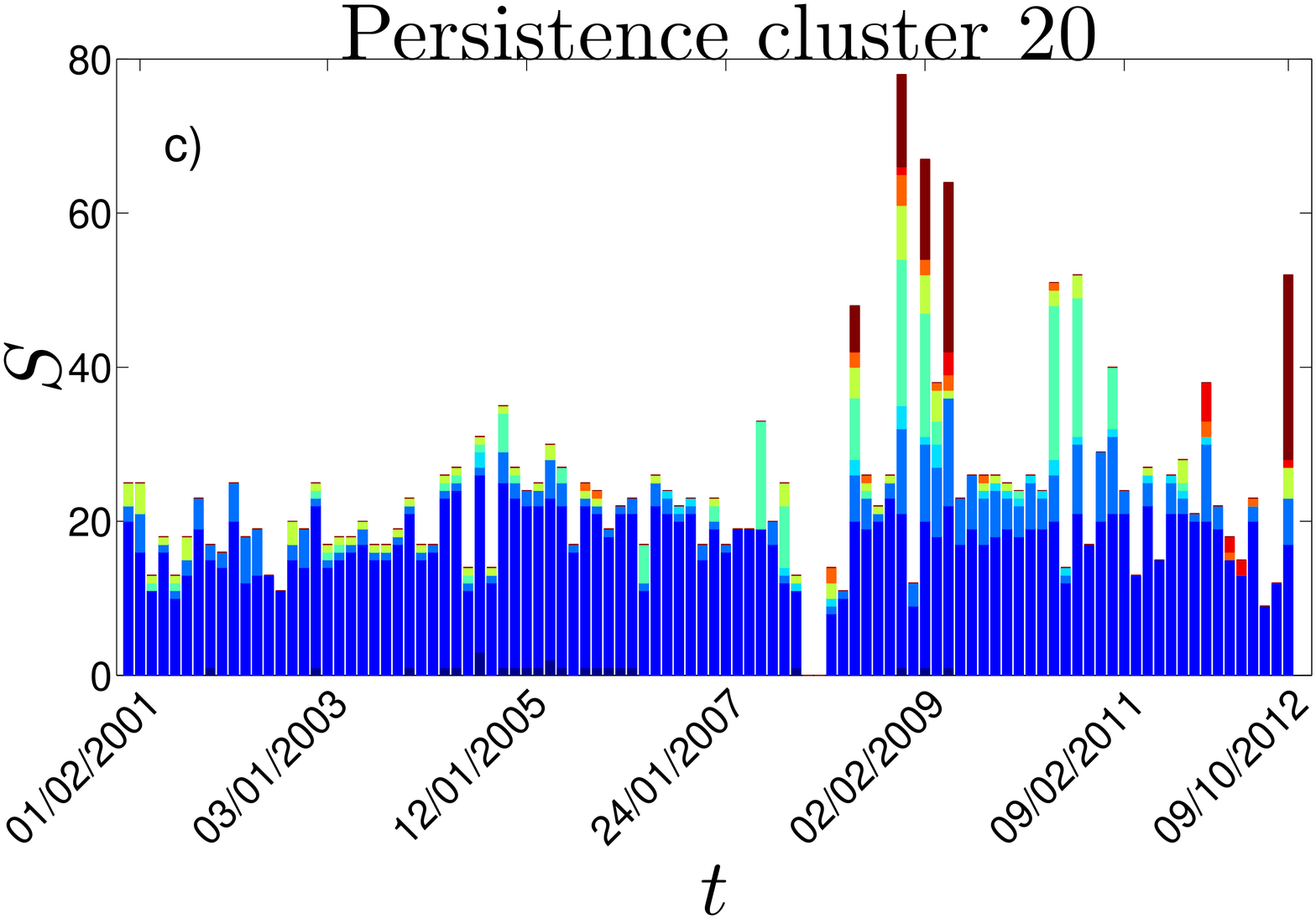}
   \includegraphics[width=0.49\textwidth]{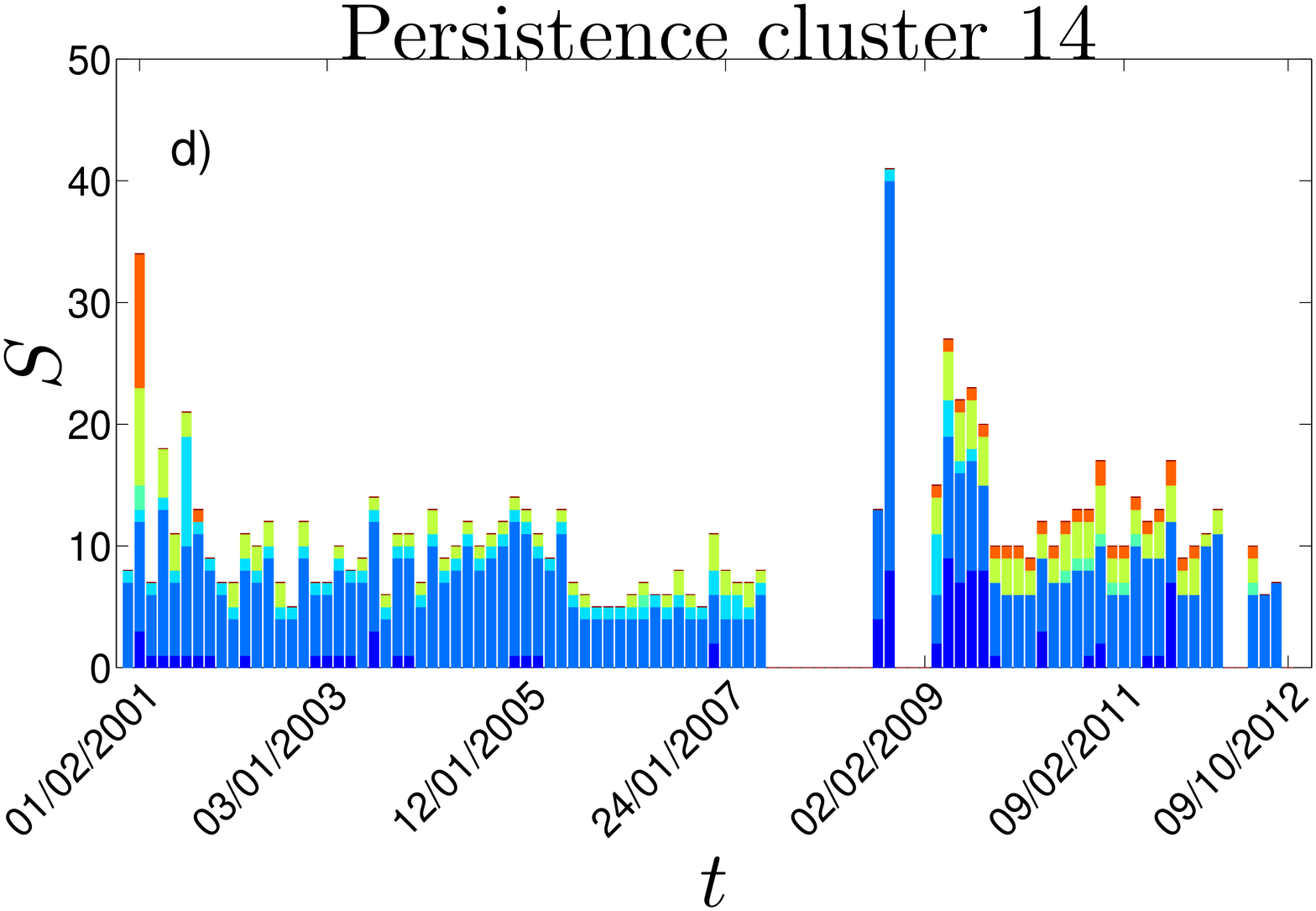}
   \includegraphics[width=0.49\textwidth]{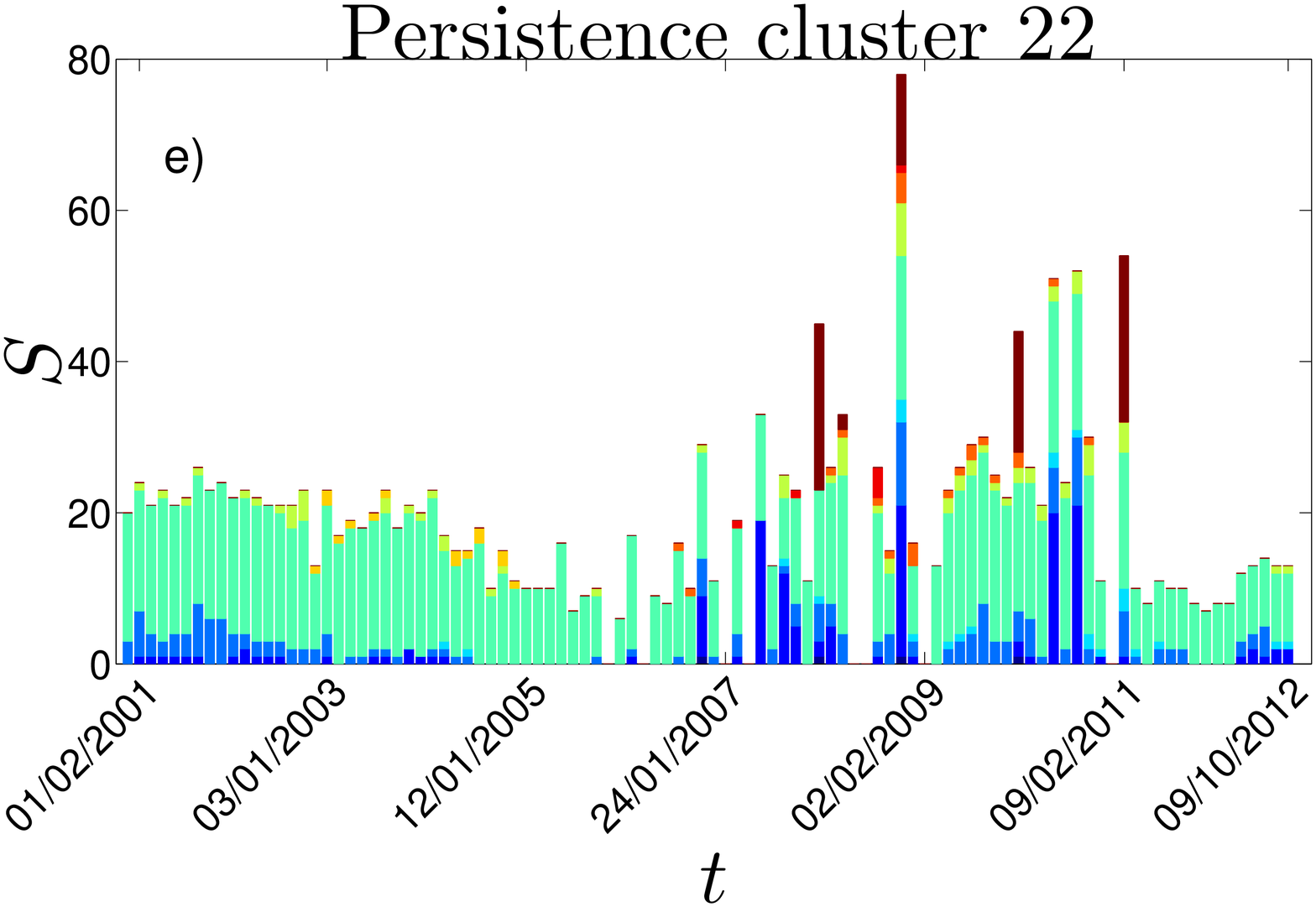}
   \includegraphics[width=0.49\textwidth]{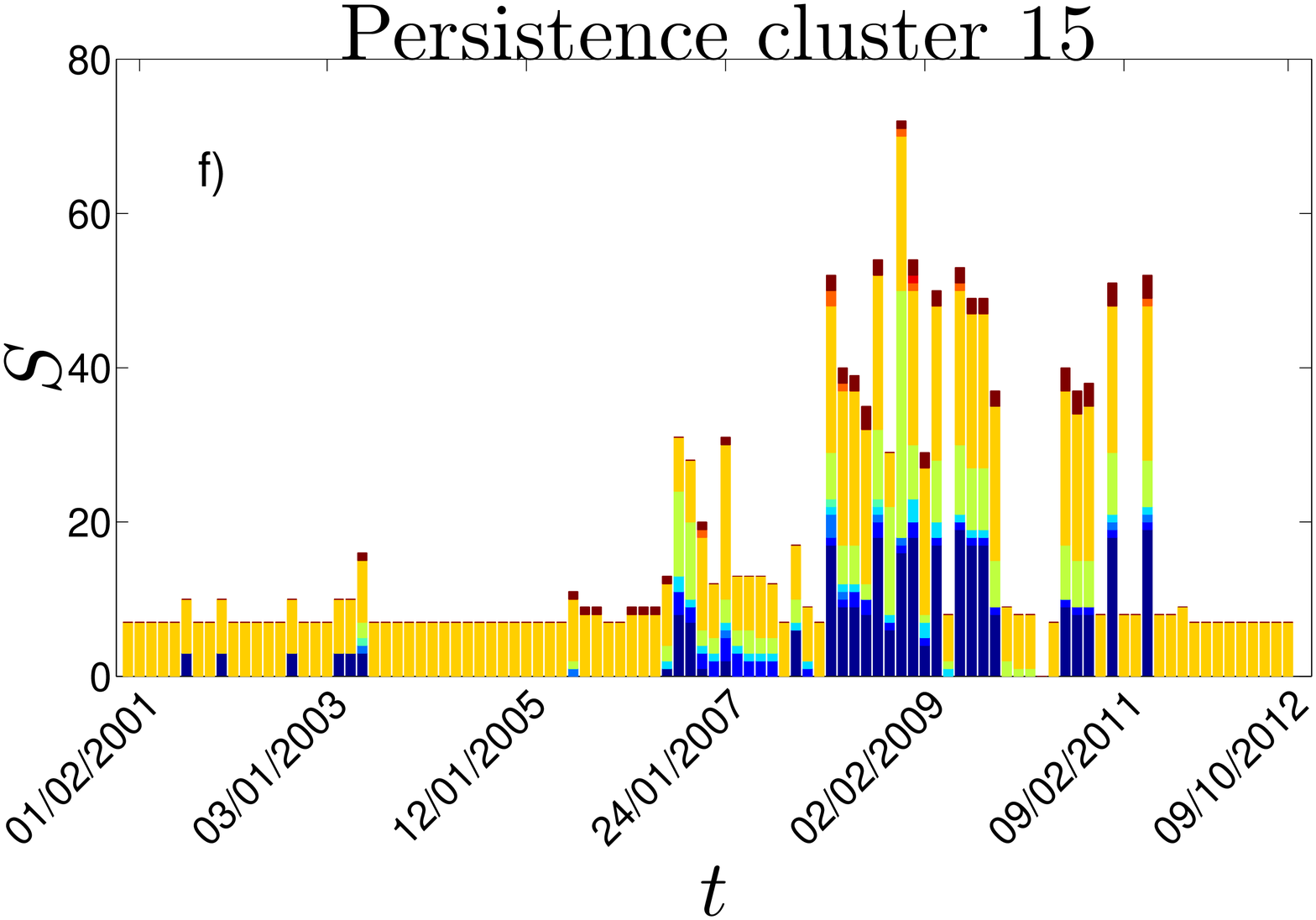}
%
%
%
%
\caption{\label{fig:cl_composition2} {\bf Clusters dynamical composition (part 2)}. 
 a) For the cluster number 1 in  Fig. \ref{fig:cl_composition} a) we have detected at each time window the correspondent ``similar''
(according to the hypergeometric test) cluster and we have plotted the composition in time. Size equal to zero corresponds to no ``similar'' cluster found. When more than one ``similar'' cluster is found only data of the largest
cluster is plotted. b)-f): same plots 
as in a), for clusters 6, 20, 14, 22 and 15 respectively. Colours refer to the legend in Fig. \ref{fig:cl_composition} a).} 
\end{figure} 

So far we have described the persistence of clusters from a global perspective, looking at the clustering as a whole. Let us here focus on the evolution of each cluster, following how their composition changes in time. It is not 
straightforward to analyse such an evolution, the main problem being the changeable nature of dynamical clusters that makes difficult to identify the successor for each cluster. Many different approaches can be adopted to
address this community 
tracking problem \citep{mst_exch_rate}. Here we use hypothesis statistical tests based on the hypergeometric distribution \citep{bipartite_net} to assess similarity between clusters at different times. In particular, if 
the number of stocks in common between two clusters is high enough to reject the null hypothesis of the test, we label the two clusters as ``similar''. Moreover we take the DBHT 
clustering calculated on the entire time window (1997-2012) as a benchmark clustering through which tracking the evolution of the dynamical clusters obtained with the moving time windows. Let us here describe the idea in more details. 

Let us call $X$ the clustering obtained on the entire time window and $Y^i$ a cluster belonging to $X$, with $i=1,..., N_{cl}$. For each cluster $Y^i$ and for each time window $T_k$ ($k=1,...,n$) we have taken the clustering 
at time $T_k$, $X_{T_k}$, and identified the cluster belonging to $X_{T_k}$ that is ``similar'' to $Y^i$ (if any). We label a cluster as ``similar'' to $Y^i$ if the number of stocks in
 common with $Y^i$ is high enough to reject the null hypothesis of the hypergeometric test \citep{musmeci_DBHT}, that considers a random overlapping between the two clusters (a detailed description of the test can be found in 
 Appendix B).
 If more than one cluster turns out to be similar, we have taken into account the largest cluster. 
 Eventually we have ended up, for each $Y^i$, with one cluster for each time window $T_k$, all of them having in common high similarity with $Y^i$. We can therefore follow their evolution in terms of number of stocks and 
 correspondent ICB industrial sectors. The threshold for the tests has been chosen equal to $0.01$, together with the conservative Bonferroni correction \citep{bipartite_net}.  
 
In Fig. \ref{fig:cl_composition} a) the composition of the DBHT clustering computed on the time window 1997-2012 is shown: for each cluster the y-axis displays its cardinality $S$ (ie, number of stocks belonging to the cluster),
 with different colours showing stocks belonging to different ICB
 industries. 
 In Figs. \ref{fig:cl_composition}  b) - f) and \ref{fig:cl_composition2}  a) - f) we plot, for the $11$ biggest clusters in $X$, the number of stocks $S$ for their similar clusters in time, together with their composition in terms of ICB industries. When for a time 
 window no similar clusters can be found we have just left empty the correspondent window. The clusters analysed are clusters $18$, $4$, $8$, $7$, $17$, $1$, $6$, $20$, $14$, $22$ and $15$. Let us here summarize the main findings:
 
 \begin{itemize}
  
  \item Overall, all the clusters in $X$ have a high persistence in time, showing a correspondent ``similar'' cluster at almost each time window. This result is even more remarkable as the persistence has been assessed in quite a 
  conservative way, ie, the hypergeometric test with Bonferroni correction. Few clusters display a limited number of gaps in their evolution (clusters 14, 15, 20 and 22), mostly in correspondence with the financial crisis. 

  \item Few clusters show a persistence in terms of industrial composition as well (it is the case of clusters 4 and, in a less extent, 8), but the majority shows a clear evolution. In particular we can distinguish quite well
  a pre-crisis and a post-crisis state, the latter characterized by a higher degree of mixing of different industries. If over the pre-crisis period we find clusters dominated by one or two industries 
  (Technology and Industrials in cluster 18, Oil \& Gas in 4 and 15, Utilities in 17, Consumer Services and Goods in 14 and 20, Financials in 6, Health Care in 22), in the crisis and post-crisis years the industries tend to mix 
  together much more, forming mixings that were not present earlier (Oil \& Gas with Basic Materials and Industrials in cluster 1 and 7, Utilities with Telecommunications and Consumer Services in 17, Financials with Consumer Goods
   and Services in 6, Health Care with Utilities and Consumer Goods in 20). This again points out the fact that the years after the crisis have seen a drop in the reliability of industries as benchmark to diversify risk.
   
   \item Apart from the pre and post-crisis dichotomy, in some cases the 2007-2008 crisis years show their own features as well. As stated above, some clusters ``disappear'' during the peak of the crisis (clusters 14, 20 and 22).
   Many others show instead several peaks in their sizes, together with a sudden increase in the number of industries: this is probably related to the merging of many clusters in few, larger clusters during the crisis. 
   
   \item The cluster containing Financial stocks (cluster 6) is worth to be analyzed further, since it seems to play a role in the outbreak of financial crisis. Indeed it shows a clear change in 2007, becoming larger and larger and 
   including an increasing number of different industries (especially Health Care, Technology and Consumer Services). This pattern is probably connected to the rising importance of the Financial industry as driving factor over 
   the outbreak of crisis. Interestingly at the end of 2008, when Lehman Brothers went bankrupt, this cluster drops suddenly to much lower sizes (although still higher than the pre-crisis values) and less mixed composition. 
   This fact suggests that the Financial industry ends playing a major role in the correlation structure from 2009 onwards.
   
 \end{itemize}
 
\section*{Discussions}
In this work we have investigated the dynamical evolution and non-stationarity of market correlation structure by means of filtered correlation networks. 
In particular, we have focused on Planar Maximally Filtered Graph (PMFG) and the clustering that its topology naturally provides by means of the Directed Bubble Hierarchical Tree (DBHT) method. 
We have measured the persistence of correlation structure by calculating similarity among clusterings at different time windows, using the Adjusted Rand Index for quantifying the similarity.

The analyses reveal that the outbreak of the 2007-2008 financial crisis marks a transition from relatively high levels of persistence to a much more unstable and changeable structure.
The minimum persistence is reached at the end of 2008 when the crisis were fully unfolded.
But the decay in persistence started already in the late 2006 well before other warning signs were detectable.
Correlation structure persistence eventually recovered in the second half of 2009 with relatively high values until the end of 2011. 
However, such a persistent structure had distinct features from the pre-crisis structure with lower relations with the industrial sectors activities. 
Notably, since the end of 2011 we are observing a new decay in persistence  which is signalling the building-up of another unfolding  change in the market structure.
This also points out that from 2007 onwards correlation matrices from historical data, both filtered and unfiltered, have became more unstable and therefore less reliable instruments for risk diversification. 
Moreover the decrease in the similarity between correlation-based clustering and industrial sector   
implies that also portfolio diversification strategies based on economic activity considerations are expected to be less effective. 
Furthermore, the analysis on the evolving industrial sector composition of each single cluster reveals that most of them display a clear change with the crisis, that overall makes them more heterogeneous in terms of industrial sectors. 
In particular, we observed that one cluster, mainly made of Financial stocks, experiences a sharp rise in its size and heterogeneity that is probably a picture of the breakdown of late 2007 financial crisis. 
This could give interesting insights in terms of early warning signals that we plan to investigate further in a future work. 

\section*{Acknowledgments}
The authors wish to thank Bloomberg for providing the data.
TDM wishes to thank the COST Action TD1210 for partially supporting this work. TA acknowledges support of the UK Economic and Social Research Council (ESRC) in funding the Systemic Risk Centre (ES/K002309/1). 

\section*{Appendix A: Adjusted Rand Index}
Following the notation of \citep{comparing_clustering}, let us call $X$
the set of $N$ objects. $Y$ is a partition into communities of $X$ or simply a clustering, that is ``a set $Y=\{Y_1,...,Y_k\}$ of non-empty disjoint subsets of $X$ such that their union equals $X$''
\citep{comparing_clustering}.
Let us say we also have another clustering $Y'$: we call ``contingency table'' the matrix $M=\{m_{ij}\}$ where 
 
 \begin{equation}
  m_{ij} \equiv |Y_i \cap {Y'}_j|,
 \end{equation}
 
 i.e. the number of objects in the intersection of clusters $Y_i$ and
 ${Y'}_j$ . 
 Let us call $a$ the number of pairs of objects that are in the same cluster both in $Y$ and in $Y'$, and $b$ the number of pairs that are in two different clusters both in $Y$ and $Y'$.
 Then the Rand Index is defined as the sum of $a$ and $b$, normalized by the total number of pairs in $X$:
 
 \begin{equation}
  \mathcal{R}(Y,Y') \equiv \frac{2(a+b)}{N(N-1)} = \sum_{i=1}^k \sum_{j=1}^l \binom{m_{ij}}{2}.
 \end{equation}

As null hypothesis associated to two independent clusterings we can assume a generalized Hypergeometric distribution, that we describe in details in the next section. 
The Adjusted Rand Index is defined as the difference between the measured Rand Index and its mean value under
the null hypothesis, normalized by the maximum that this difference can reach:

\begin{equation}
\label{eq:ari}
 \mathcal{R}_{adj}(Y,Y') \equiv \frac{\sum_{i=1}^k \sum_{j=1}^l \binom{m_{ij}}{2}-t_3}{\frac{1}{2}(t_1+t_2)-t_3},
\end{equation}

where

\begin{equation}
 t_1=\sum_i^k \binom{|Y_i|}{2} ~~ , ~~ t_2=\sum_j^l \binom{|{Y'}_j|}{2} ~~,~~ t_3=\frac{2t_1t_2}{N(N-1)}.
\end{equation}

It turns out that $\mathcal{R}_{adj} \in [-1, 1]$ , with $1$ correspondent to the case of identical clusterings and $0$ to two completely uncorrelated clusterings. Negative values instead show anti-correlation between $Y$ and $Y'$
 (that is, the number of pairs classified in the same way by $Y$ and $Y'$ is less even than what expected assuming a random overlapping between the two clusterings). 
 
\subsection*{Appendix B: Hypergeometric Test}
\label{appendix:hypergeometric_test}
Following the notation we have used in Section ``Clusters composition evolution'', let us call $Y_i$ a generic cluster belonging to the clustering calculated on the entire time window. Let $Y'_j$ be the cluster 
from clustering $X_{T_k}$ at time window $T_k$ with which we want to compare $Y_i$ in order to answer the question ``is the number of stocks belonging to both $Y_i$ and $Y'_j$ sensitively higher than what expected by a random overlapping?''.
The question can be translated into a statistical one-tail hypothesis test where the null hypothesis is the Hypergeometric distribution. Say $k$ is the number of stocks in common between $Y_i$ and  $Y'_j$, 
whereas $|Y_i|$ , $|Y'_j|$ are the
 cardinalities of the two clusters; then  
the Hypergeometric distribution reads \citep{bipartite_net}:

\begin{equation}
\label{eq:hypog_test}
 P(X=k) ~=~ \frac{{|Y'_j|\choose k}{N-|Y'_j|\choose|Y_i|- k }}{{N\choose|Y_i| }} 
\end{equation}
 
 This distribution is consistent with a scenario where the overlapping between the two clusters is due purely to chance. For this reason it is a suitable null hypothesis for testing similarity between clusters.
 If $P(X=k)$ so calculated is less than the significance level, then the test is rejected and we conclude that the cluster $Y_i$
 over-expresses the cluster $Y'_j$ and they are therefore similar. The significance level of each test performed is $1\%$, together with the Bonferroni correction for multiple tests \citep{bipartite_net}. 

\bibliographystyle{plainnat}
\bibliography{bibl_sample.bib}

\end{document}